\documentclass[11pt]{article}

\usepackage[preprint]{acl}

\usepackage{times}
\usepackage{latexsym}

\usepackage[T1]{fontenc}

\usepackage[utf8]{inputenc}

\usepackage{microtype}
\usepackage{enumitem}
\usepackage{inconsolata}

\usepackage{graphicx}

\usepackage[ruled,vlined]{algorithm2e}
\usepackage{subcaption}
\usepackage{graphicx}
\usepackage{booktabs}
\usepackage{multirow}
\usepackage{makecell}
\usepackage{adjustbox}
\usepackage[table]{xcolor}
\usepackage{setspace}
\usepackage[most]{tcolorbox}
\definecolor{agentcolor}{RGB}{255, 255, 255}
\newtcolorbox{dialoguebox}[3][]{
    breakable, 
    enhanced,
    before skip=3pt, 
    after skip=3pt,
    left=3pt, right=3pt, top=3pt, bottom=3pt,
    colframe=#2!60!black, 
    colback=#2,
    title={\textbf{#3}}, 
    fonttitle=\sffamily\tiny,
    fontupper=\footnotesize, 
    before upper={\setstretch{1.05}}, 
    arc=2pt,
    boxrule=1.5pt,
    parbox=false, 
    #1
}
\newcommand{\gain}[1]{\textcolor{teal}{\scriptsize(#1)}}

\title{GUITestScape: Towards Open-set Evaluation on Exploratory GUI Testing}

\author{
  \textbf{Xiaoyi Chen\textsuperscript{1\large$\ast$}}, 
  \textbf{Yifei Gao\textsuperscript{1}}, 
  \textbf{Yang Xu\textsuperscript{1}}, 
  \textbf{Zhenlong Wang\textsuperscript{1}}\\
  \textbf{Jian Yan\textsuperscript{1}}, 
  \textbf{Jiming An\textsuperscript{1}}, 
  \textbf{Xingxing Song\textsuperscript{1}}, 
  \textbf{Liqun Deng\textsuperscript{2}}\\
  \textbf{Xiaojun Meng\textsuperscript{2}}, 
  \textbf{Jiansheng Wei\textsuperscript{2}}, 
  \textbf{Yi Zhang\textsuperscript{3}}, 
  \textbf{Jitao Sang\textsuperscript{1$\dagger$}}\\
  \\
  \textsuperscript{1}Beijing Jiaotong University,  
  \textsuperscript{2}Huawei Noah's Ark Lab, 
  \textsuperscript{3}Independent Researcher
}

\begin{document}
\maketitle

\begingroup
\renewcommand{\thefootnote}{\fnsymbol{footnote}}
\footnotetext[1]{Part of this work was conducted by the first author at Huawei Noah's Ark Lab, with support from the lab.}
\footnotetext[2]{Corresponding author: \href{mailto:jtsang@bjtu.edu.cn}{jtsang@bjtu.edu.cn}}
\endgroup

\begin{abstract}

  Exploratory GUI testing is a particularly demanding setting for MLLM agents: without predefined test scripts, an agent must autonomously navigate an application and discover defects through its own interaction. However, current evaluation falls short on two fronts. First, existing benchmarks focus almost exclusively on interaction defects, leaving display defects outside the evaluation frame. Second, evaluation protocols are bound to predefined defect annotations, collapsing the testing process into a single end-state judgment that conflates qualitatively distinct failure modes. To address these challenges, we present \textbf{GUITestScape}, an interactive benchmark covering 61 real-world Android applications and 508 preset defects spanning interaction and display types, and introduce \textbf{GUIJudge}, an open-set evaluator that decomposes an agent's testing trajectory into independently diagnosable capabilities. Experimental results demonstrate that GUIJudge achieves reliable process-aware evaluation beyond predefined annotations, substantially outperforming all baselines. Benchmarking on GUITestScape further reveals that detection remains the critical bottleneck for existing models across both defect types, and that integrating GUIJudge's verifiers into existing agents significantly boosts their detection performance without retraining.

\end{abstract}

\section{Introduction}

Multimodal Large Language Models have substantially strengthened GUI agents, extending their reach from single-step grounding to long-horizon tasks~\cite{qin2025uitarspioneeringautomatedgui,rawles2025androidworlddynamicbenchmarkingenvironment,xie2024osworld} and practical applications such as software quality assurance~\cite{zhao2024gui}. Among these, exploratory GUI testing stands out as particularly demanding: without predefined test scripts, an agent must autonomously navigate an application and discover defects through its own interaction~\cite{gao2026guitester}. Yet current evaluation falls short of measuring these capabilities faithfully.

\begin{figure*}[t]
    \centering
    \begin{subfigure}[t]{0.34\textwidth}
        \centering
        \raisebox{1mm}{\includegraphics[width=\linewidth]{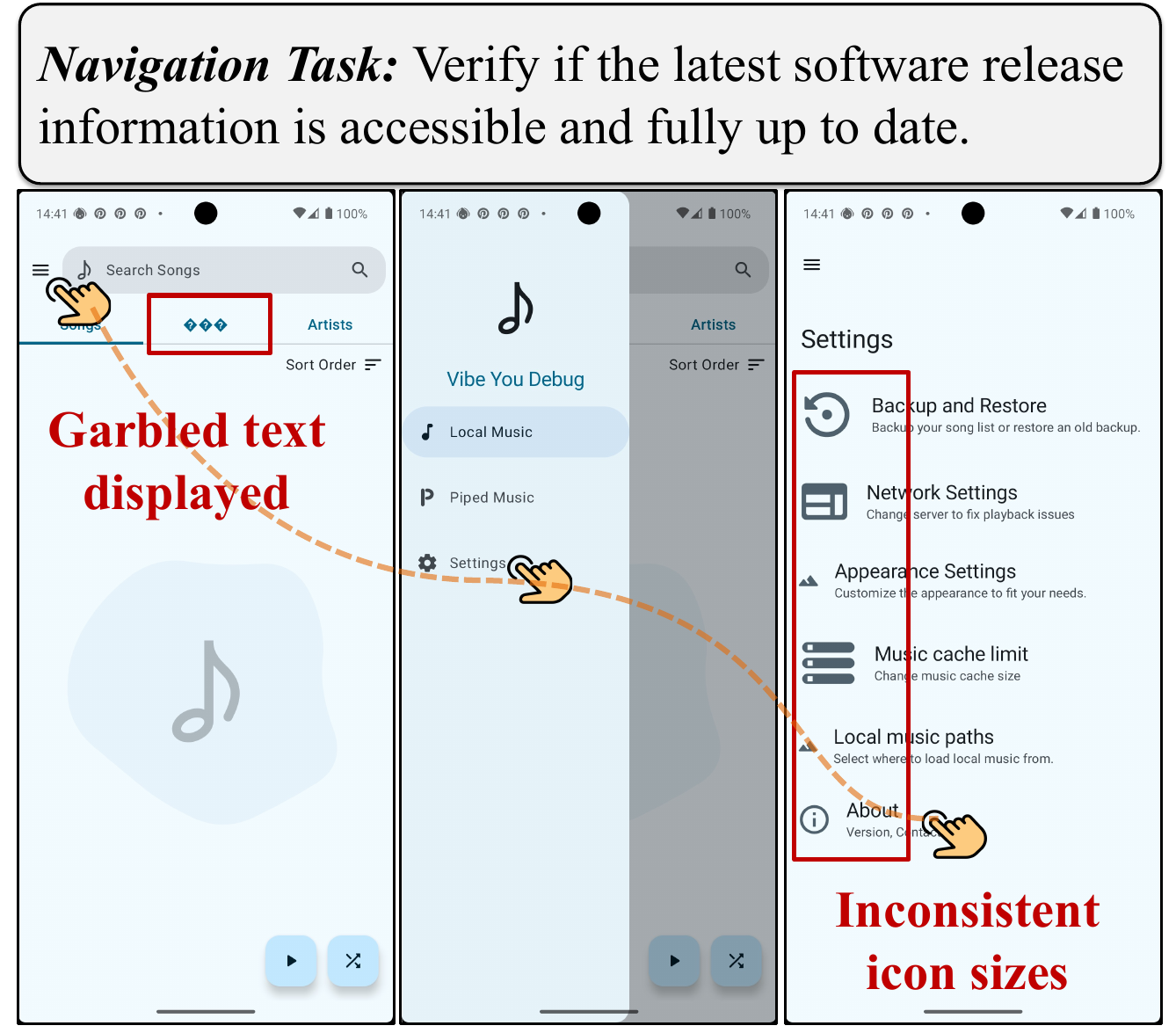}}
        \caption{Display defects in one test case.}
        \label{fig:intro_display_defects}
    \end{subfigure}
    \hfill
    \begin{subfigure}[t]{0.62\textwidth}
        \centering
        \includegraphics[width=\linewidth]{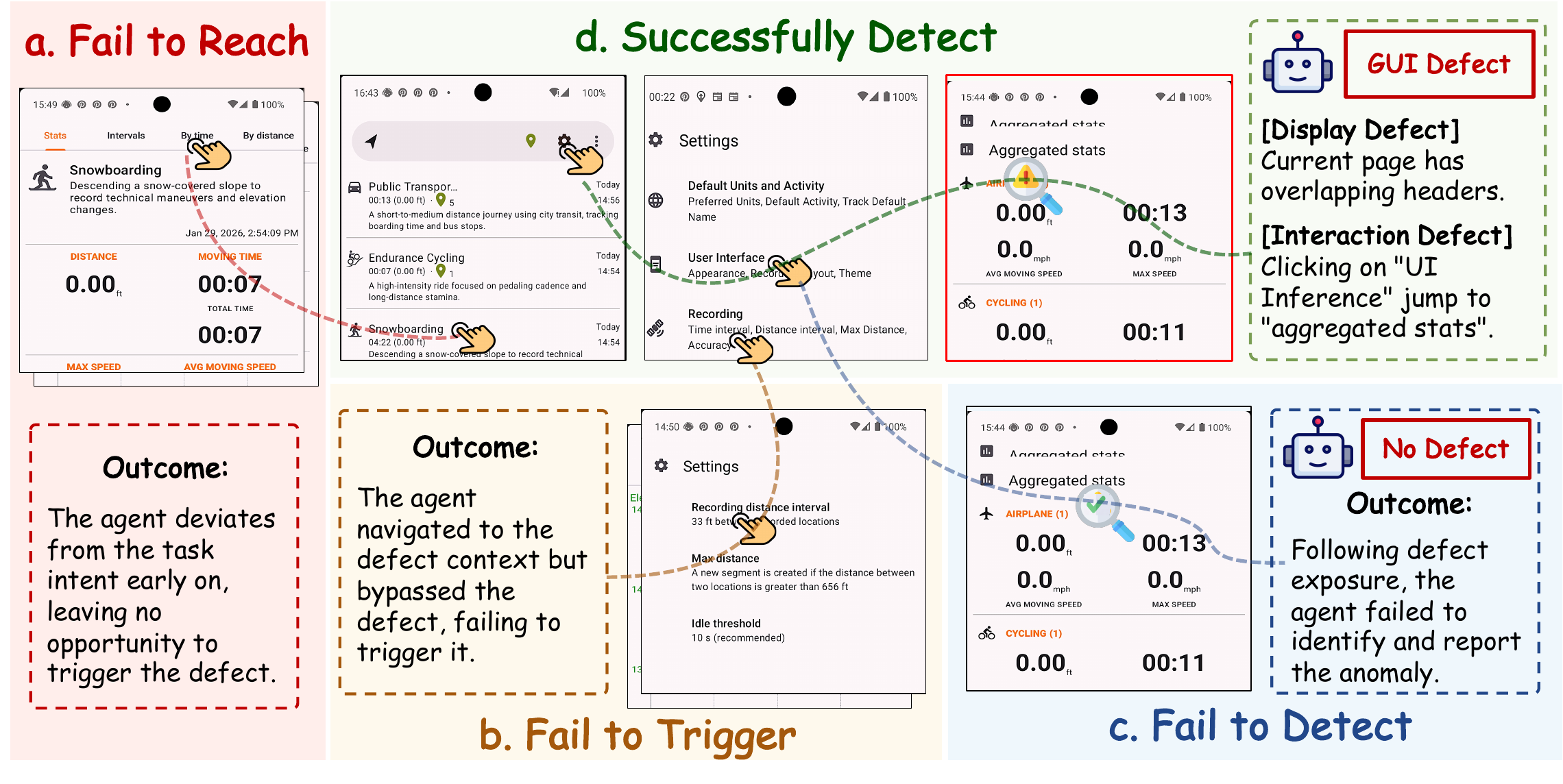}
        \caption{Different outcomes in exploratory GUI testing.}
        \label{fig:intro_failure_modes}
    \end{subfigure}
    \caption{
    Motivating examples for GUITestScape and GUIJudge.
    (a) A realistic testing process exposes multiple display defects, such as garbled text and inconsistent icon sizes.
    (b) The same exploratory testing process may lead to different outcomes, including failures at Reach, Trigger, and Detect, as well as successful defect detection.
    }
    \label{fig:intro_motivation}
\end{figure*}

We identify two fundamental challenges: (1) \textbf{Display-defect blindness}. Current benchmarks \cite{gao2026guitester,ahmed2026specops} focus almost exclusively on interaction defects, overlooking display defects such as rendering errors, missing elements, and layout misalignment~\cite{liu2020owl,liu2022nighthawk}. These defects often have the most immediate impact on user experience, yet detecting them demands open-ended visual anomaly detection, a capability that interaction-oriented evaluation never probes. (2) \textbf{Closed-set evaluation}. Existing protocols verify agent reports against predefined defect annotations, which introduces two limitations. First, the evaluation cannot scale to unlabeled defects, since any anomaly outside the annotated set has no reference for verification. Second, scoring collapses the entire testing process into a single end-state judgment~\cite{he2024webvoyager}, conflating qualitatively distinct failure modes. An agent may never reach the defect scenario, may reach it but fail to trigger the defect, or may trigger the anomaly yet fail to recognize it. Consequently, agents with very different capability profiles are flattened to indistinguishable scores. Figure~\ref{fig:intro_motivation} illustrates these challenges with two examples: a realistic interaction case where display defects appear repeatedly during testing, and an exploratory testing process where agents exhibit different failure modes before successful defect detection.

To address these challenges, we present \textbf{GUITestScape}, an interactive benchmark covering 61 real-world Android applications and 508 preset defects spanning interaction and display types. Each defect is instantiated into a structured evaluation case along three dimensions: a resettable scenario that ensures run-to-run consistency, a scenario-level test basis that captures app-specific correctness semantics, and multiple navigation tasks that simulate how testers might approach the same defect through different interaction patterns. We further introduce \textbf{GUIJudge}, an open-set evaluator that decomposes an agent's testing trajectory into three independently diagnosable capabilities: \textit{Reaching}, \textit{Triggering}, and \textit{Detection}, disentangling failure modes that prior end-state scoring conflates.

Experimental results demonstrate that GUIJudge achieves reliable process-aware evaluation, outperforming all baseline evaluators with 97.12\% Recall at the Detect stage. GUITestScape further reveals that detection remains the critical bottleneck for existing models, with the best-performing model reaching only 57.02\% Recall. This bottleneck can be effectively addressed by integrating GUIJudge's verifiers into existing agents without retraining. Our contributions include: 

\begin{itemize}[leftmargin=1em, itemsep=-2pt]
\item We present \textbf{GUITestScape}, an interactive benchmark covering 61 real-world Android applications and 508 preset defects across both interaction and display types.
\item We propose a \textbf{three-stage capability decomposition} that disentangles exploratory GUI testing into Reaching, Triggering, and Detection as independently diagnosable dimensions.
\item We introduce \textbf{GUIJudge}, an open-set evaluator that enables process-aware evaluation beyond predefined defect annotations.
\end{itemize}

\section{Related Works}




\subsection{Automated GUI Testing}

Automated GUI testing has long relied on random, heuristic, or UI-guided exploration. Tools such as Monkey~\cite{7372031} and DroidBot~\cite{10.1109/ICSE-C.2017.8} traverse application interfaces, but they lack awareness of app intent and defect semantics. Recent work has introduced MLLM-based agents into GUI testing. ScenGen~\cite{yu2026scenarioguidedllmbasedmobileapp}, VisionDroid~\cite{liu2024seeingbelievingvisiondrivennoncrash}, GUITester~\cite{gao2026guitester}, and SpecOps~\cite{ahmed2026specops} extend GUI testing toward intent-aware and defect-oriented exploration.

\subsection{Benchmarks for GUI Testing}

Existing GUI benchmarks such as AndroidWorld~\cite{rawles2025androidworlddynamicbenchmarkingenvironment} and OSWorld~\cite{xie2024osworld} mainly evaluate task completion. GUI testing benchmarks instead focus on testing capabilities: GUI Testing Arena~\cite{zhao2024gui} structures evaluation around intent generation, task execution, and defect detection; GUITester~\cite{gao2026guitester} introduces GUITestBench for exploratory defect discovery; and ScenGen~\cite{yu2026scenarioguidedllmbasedmobileapp} emphasizes scenario-based test construction.

\subsection{GUI Trajectory Evaluation}

Evaluating exploratory GUI trajectories is challenging because valid trigger paths are not unique and defect evidence may appear only in short trajectory intervals. Rule-based methods, including those used in GUITester~\cite{gao2026guitester} and WebArena~\cite{ICLR2024_4410c071}, rely on predefined checks or task-specific references, limiting generalization beyond closed-set settings. LLM-as-Judge methods - have also been explored in agent evaluation, including Mind2Web~\cite{deng2023mind2web}, AgentRewardBench~\cite{lù2025agentrewardbenchevaluatingautomaticevaluations}, and Beyond Binary Rewards~\cite{damani2025binaryrewardstraininglms}. However, holistic judgments over long trajectories provide limited diagnosis of where the testing process fails.

\section{GUITestScape}

\begin{figure*}[t]

    \centering

    \includegraphics[width=\textwidth]{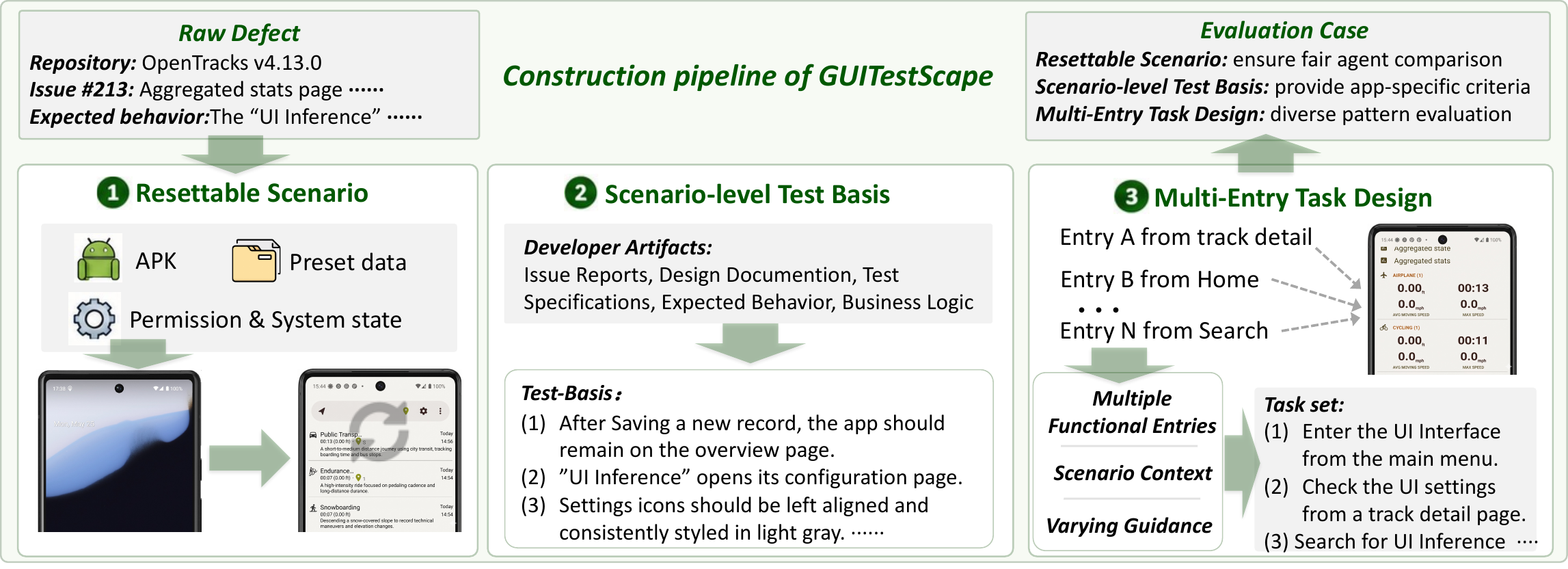}
    \caption{Construction pipeline of GUITestScape. Collected defects are instantiated into an evaluation case with a resettable scenario, a scenario-level test basis, and a multi-entry task set.}
    \label{fig:guitestscape_pipeline}

\end{figure*}

This section presents GUITestScape, an interactive benchmark for evaluating agents in exploratory GUI testing. Figure~\ref{fig:guitestscape_pipeline} provides an overview of its construction pipeline. We first describe the collection and categorization of real-world defects (\S3.1). We then detail the benchmark composition, where resettable scenarios provide reproducible execution environments containing preset defects, each paired with a scenario-level test basis and multiple navigation tasks (\S3.2). Finally, we report the scale and distribution of GUITestScape (\S3.3).

\subsection{Defect Collection and Categorization}

We collect GUI defects from public issue trackers and bug reports of real-world Android applications. Building on the defect taxonomy of GTArena~\cite{zhao2024gui}, we organize the collected defects into two categories, display defects and interaction defects.



\noindent\textbf{Display Defects} are visual abnormalities in the rendered interface. Detecting them requires open-ended visual anomaly perception from visual evidence alone. We further divide them into two fault modes. (1) \textit{Content Rendering} defects (CR) concern the abnormal presentation of textual or visual content, such as broken icons or garbled text. (2) \textit{Element Layout} defects (EL) concern abnormalities in spatial organization, such as overlapping elements or alignment errors.

\noindent\textbf{Interaction Defects} are anomalies arising from inconsistencies between an agent's actions and the application's responses, requiring reasoning over the resulting interaction trajectory rather than visual evidence alone. We adopt three fault modes from GUITester~\cite{gao2026guitester}. (1) \textit{Operation No Response} (ONR), where an interaction yields no observable feedback. (2) \textit{Unexpected Task Result} (UTR), where the outcome of a multi-step operation deviates from its intended effect. (3) \textit{Navigation Logic Error} (NLE), where flawed transitions lead to incorrect navigation flow.

\subsection{Benchmark Composition}


A benchmark requires a dynamic case in which an agent can actually encounter, trigger, and judge a defect. We instantiate the collected defects within a set of structured evaluation cases, each characterized along three dimensions:
(1) a resettable scenario (\S3.2.1) that restores the same starting environment before every run; 
(2) a scenario-level test basis (\S3.2.2) that provides app-specific correctness criteria;
and (3) multiple navigation tasks (\S3.2.3) that simulate the diverse interaction patterns of real testers.

\subsubsection{Resettable Scenario}



Whether a defect remains reproducible depends on the application's runtime context~\cite{romano2021empirical}, and uncontrolled drift in this context makes evaluation unstable across runs. We therefore wrap each defect in a resettable evaluation scenario that fixes its prerequisites for triggering. Beyond the target application, we specify the defect-relevant initial conditions under which testing begins, including pre-populated application data and required system-side settings. These conditions are restored before every evaluation, so that different agents are tested from an identical starting point when probing the same defect. Consequently, evaluation results reflect the agent's testing competence rather than environmental contingency.

\subsubsection{Scenario-level Test Basis}


In exploratory GUI testing, a GUI defect is identified when the observed GUI behavior deviates from the expected behavior. However, generic interaction knowledge alone is insufficient to determine the expected behavior across different applications. For example, when a feedback form returns the user to the home screen after submission, a tester familiar with the app's design rationale recognizes this as the intended streamlined flow, whereas the same transition in an app that should keep the user on the editable page would be flagged as a navigation logic error. Testers therefore interpret interaction results according to a test basis~\cite{schieferdecker2025navigatinggrowingfieldresearch,istqb_glossary}, which captures the design specifications and business logic of the target application.

Following the ISTQB definition~\cite{istqb_glossary}, for each resettable scenario in GUITestScape, we construct a scenario-level test basis from developer artifacts of the application under test, structuring the design specifications and business logic for test analysis. The resulting test basis consists of app-specific correctness criteria without prescribing predefined defect instances, execution states, or a unique action trajectory, thereby accommodating diverse interaction patterns. Each criterion specifies a precondition, the interaction or event under which the expected behavior should hold, and the observable evidence for judging whether the resulting behavior conforms to that expectation.

\subsubsection{Multi-Entry Task Design}


Human testers exploring an application rarely approach a defect through a single fixed path; they typically vary entry points and interaction sequences to rule out coincidental triggers and to expose the defect under different operational conditions. To preserve the exploratory nature of testing, GUITestScape constructs multiple navigation tasks that approach defect-relevant functionality from different functional entry points and with varying levels of guidance, without revealing the defect, expected behavior, or a prescribed interaction path. The defect therefore remains the evaluation target, while the path becomes a controlled source of variation, allowing us to assess whether an agent's testing competence generalizes across interaction patterns rather than relying on a single predefined route.

\subsection{Benchmark Statistics}

\begin{figure}[t]
    \centering
    \includegraphics[width=\columnwidth]{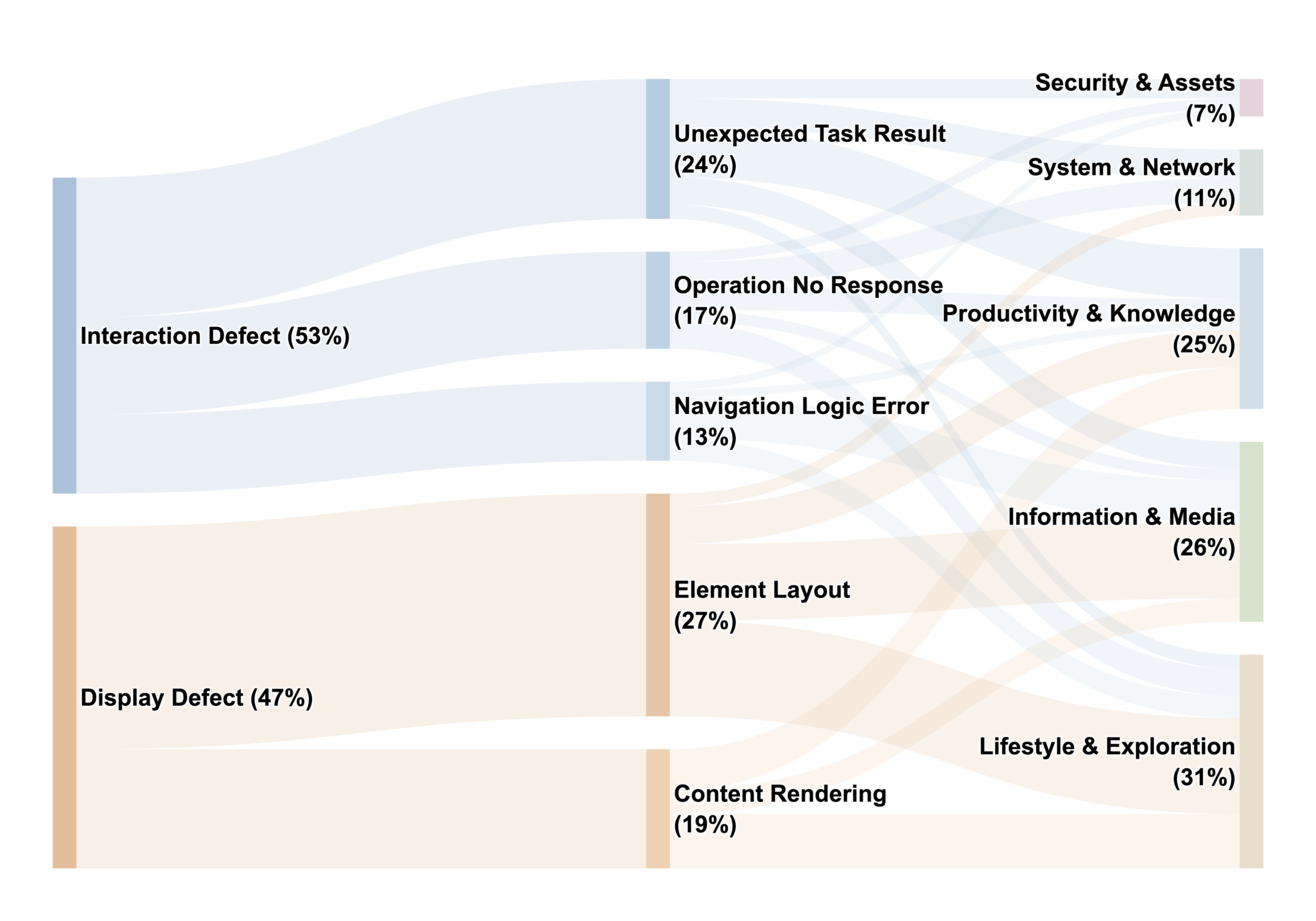}
    \caption{
    Hierarchical distribution of preset defects across defect types, fault modes, and application categories in GUITestScape.
    }
    \label{fig:benchmark_statistics}
\end{figure}

GUITestScape collects 508 preset defects from 61 real-world Android applications across 5 major categories. We consolidate these defects into 82 resettable scenarios, annotate each with scenario-level test basis, and expand them into 384 navigation tasks. These quantities are not in one-to-one correspondence: each scenario may contain multiple preset defects and supports multiple navigation tasks, while the same defect may be exposed by multiple tasks. Figure~\ref{fig:benchmark_statistics} shows the distribution across defect types and application categories, with defects nearly evenly split between interaction (53\%) and display (47\%) types.

\section{GUIJudge}

\begin{figure}[t]

    \centering
    \includegraphics[width=\columnwidth]{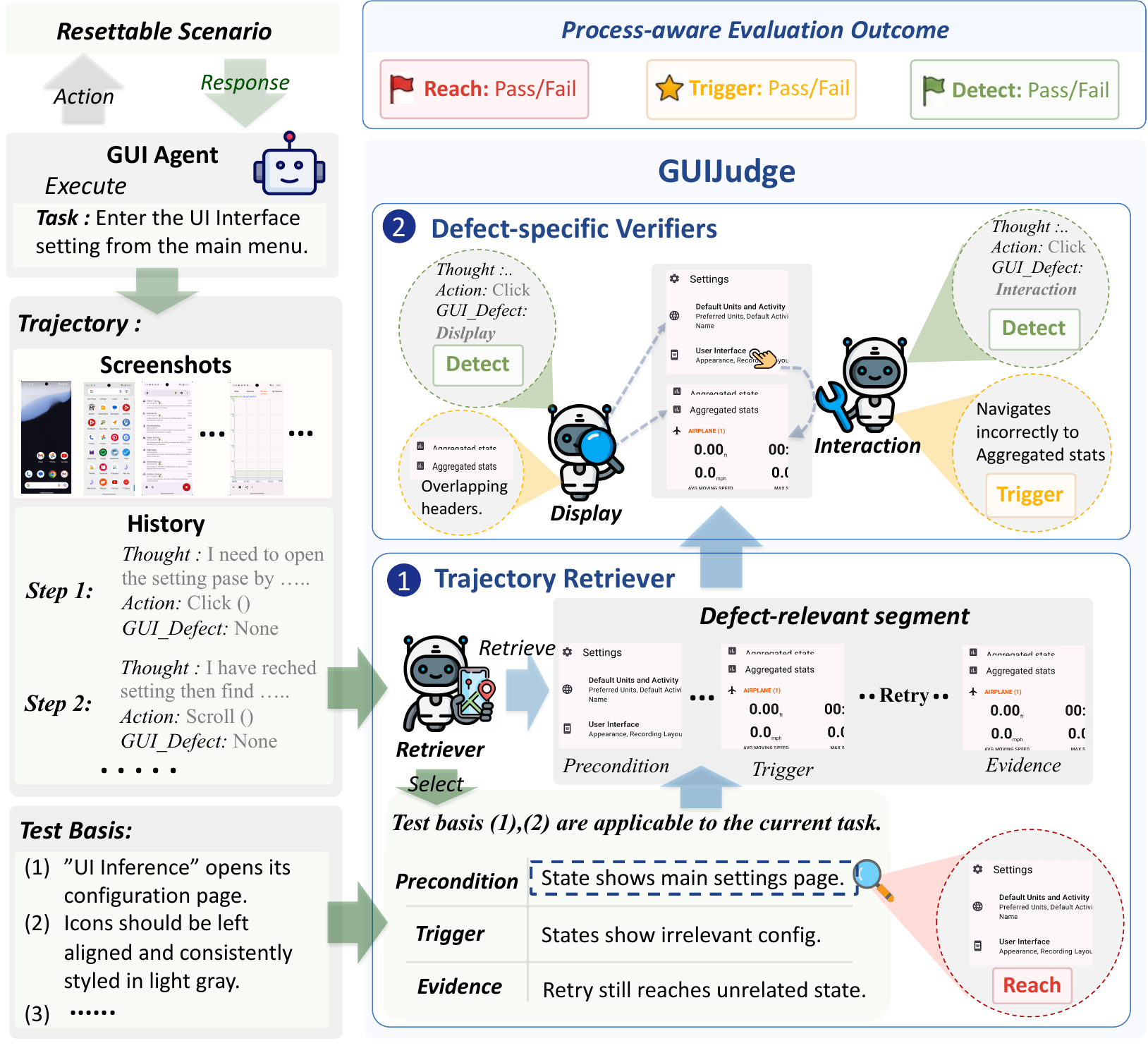}
    \caption{GUIJudge Workflow. Given an agent trajectory and scenario-level test basis, GUIJudge first selects the applicable correctness criterion from the test basis and retrieves defect-relevant segments, then applies two defect-specific verifiers to produce process-aware judgments across Reach, Trigger, and Detect.}
    \label{fig:guijudge_workflow}
\end{figure}

Building on GUITestScape, we introduce GUIJudge, a process-aware evaluator for exploratory GUI testing. We first define its open-set evaluation protocol (\S4.1) and three evaluation targets (\S4.2), then present the two-stage pipeline comprising a trajectory retriever and defect-specific verifiers (\S4.3). Figure~\ref{fig:guijudge_workflow} provides an overview.

\subsection{Open-set Evaluation Protocol}

GUIJudge follows an open-set evaluation protocol: given the scenario-level test basis, it judges observed trajectories and reports against the applicable correctness criterion rather than matching them to a finite set of predefined trajectories, defect instances, or reports. The test basis therefore defines the correctness criterion without enumerating the possible evaluation outcomes.



\subsection{Evaluation Targets}
Defect discovery unfolds in three stages: an agent must first reach the defect-relevant context, then trigger the defect through interaction, and finally recognize the resulting anomaly. A breakdown at any stage halts this progression and leaves the defect unreported. Yet existing end-state evaluation collapses the entire progression into a single verdict, masking the capability gap that actually caused the failure. GUIJudge therefore evaluates each stage as an independent dimension.

\begin{itemize}[leftmargin=1em,itemsep=-2pt]
    \item \textit{\textbf{Reaching}} evaluates whether an agent correctly interprets the testing task and navigates to the precondition state required for defect exposure.
    \item \textit{\textbf{Triggering}} evaluates whether an agent, having reached the defect-relevant context, performs the interaction that surfaces the defect. This dimension applies only to interaction defects.
    \item \textit{\textbf{Detection}} evaluates whether an agent, once the defect is exposed, recognizes the anomaly and reports it as a GUI defect.
\end{itemize}

\subsection{Two-Stage Evaluation Pipeline}

Evaluating an agent's testing trajectory faces two practical obstacles. First, exploratory trajectories interleave defect-relevant moments with extensive irrelevant navigation, making end-to-end judgment prone to evidence dilution. Second, display and interaction defects manifest through fundamentally different forms of evidence, precluding a single unified verification procedure. These two challenges motivate the design of the trajectory retriever (\S4.3.1) and the defect-specific verifiers (\S4.3.2).

\subsubsection{Trajectory Retriever}




To reduce trajectory noise caused by defect-irrelevant steps, GUIJudge first selects the applicable correctness criterion from the test basis and retrieves candidate segments according to its precondition, relevant interaction or event, and observable evidence. If the required precondition is identified, GUIJudge considers the agent to have reached the target context; otherwise, the defect discovery process fails before exposure. The retrieved segments are not treated as final defect judgments. Instead, they are passed to defect-specific verifiers, which further judge whether the defect is triggered and whether the agent reports the anomaly correctly.

\subsubsection{Defect-specific Verifiers}

Since display and interaction defects rely on different evidence forms, GUIJudge applies defect-specific verifiers to the retrieved defect-relevant segments under the selected correctness criterion. For display defects, the verifier examines whether the observed GUI states violate the selected correctness criterion. For interaction defects, the verifier examines action-state transitions and checks whether the observed response is consistent with the task expectation and the correctness criterion. The verifiers judge \textit{Trigger} by determining whether the target defect is actually exposed, and judge \textit{Detect} by checking whether the agent's report is consistent with the verified defect evidence.

\begin{table*}[t]
\centering
\small
\setlength{\tabcolsep}{4.0pt}
\renewcommand{\arraystretch}{1.12}

\begin{adjustbox}{max width=\textwidth}
\begin{tabular}{p{2.8cm}p{1.0cm}cccccccccccc}
\toprule
\multirow{3}{*}{\textbf{Method}} & \multirow{3}{*}{\textbf{Stage}}
& \multicolumn{4}{c}{\textbf{Display Defects}}
& \multicolumn{6}{c}{\textbf{Interaction Defects}}
& \multicolumn{2}{c}{\textbf{Overall}} \\
\cmidrule(lr){3-6} \cmidrule(lr){7-12} \cmidrule(lr){13-14}
& 
& \multicolumn{2}{c}{\textbf{CR}}
& \multicolumn{2}{c}{\textbf{EL}}
& \multicolumn{2}{c}{\textbf{NLE}}
& \multicolumn{2}{c}{\textbf{ONR}}
& \multicolumn{2}{c}{\textbf{UTR}}
& \multicolumn{2}{c}{\textbf{Overall}} \\
\cmidrule(lr){3-4} \cmidrule(lr){5-6}
\cmidrule(lr){7-8} \cmidrule(lr){9-10}
\cmidrule(lr){11-12} \cmidrule(lr){13-14}
& 
& Recall$\uparrow$ & F1$\uparrow$
& Recall$\uparrow$ & F1$\uparrow$
& Recall$\uparrow$ & F1$\uparrow$
& Recall$\uparrow$ & F1$\uparrow$
& Recall$\uparrow$ & F1$\uparrow$
& Recall$\uparrow$ & F1$\uparrow$ \\
\midrule

\rowcolor[gray]{0.94}
\multicolumn{14}{l}{\textit{GUIJudge (Process-Aware Performance)}} \\

GUIJudge & Reach
& 100.00 & 94.34
& 100.00 & 100.00
& 100.00 & 98.25
& 100.00 & 100.00
& 100.00 & 97.03
& 100.00 & 97.87 \\

GUIJudge & Trigger
& 100.00 & 94.34
& 100.00 & 100.00
& 100.00 & 96.77
& 100.00 & 100.00
& 100.00 & 94.92
& 100.00 & 97.12 \\

GUIJudge & Detect
& \textbf{100.00} & \textbf{89.36}
& \textbf{96.00} & \textbf{92.31}
& \textbf{100.00} & \textbf{96.77}
& \textbf{100.00} & \textbf{100.00}
& \textbf{92.31} & \textbf{94.12}
& \textbf{97.12} & \textbf{93.95} \\

\midrule

\rowcolor[gray]{0.94}
\multicolumn{14}{l}{\textit{Ablation of  GUIJudge Design}} \\

w/o Retrieval & Detect
& 85.71 & 72.39
& 70.67 & 66.69
& 75.56 & 78.02
& 94.12 & 90.39
& 65.38 & 69.58
& 76.92 & 74.23 \\

w/o Typed Verifiers & Detect
& 95.24 & 90.91
& 84.00 & 89.37
& 80.00 & 85.71
& 82.35 & 90.32
& 65.38 & 79.07
& 80.77 & 87.05 \\

\midrule

\rowcolor[gray]{0.94}
\multicolumn{14}{l}{\textit{Baseline Evaluators (Detect Only)}} \\

GUI-Tester & Detect
& -- & --
& -- & --
& \underline{86.67} & 89.66
& 62.51 & 76.92
& \underline{72.58} & \underline{78.26}
& \underline{73.21} & \underline{81.19} \\

Rule-based & Detect
& -- & --
& -- & --
& \underline{86.67} & \underline{92.68}
& 68.75 & 81.48
& 60.13 & 71.43
& 69.64 & 80.41 \\

Claude-4.6-opus & Detect
& 66.67 & 71.67
& 14.67 & 21.94
& \underline{86.67} & 78.79
& \underline{92.16} & \underline{83.15}
& 62.82 & 62.41
& 60.26 & 64.14 \\

GPT-5.4 & Detect
& 53.97 & 61.24
& 24.12 & 35.91
& 77.78 & 76.91
& 60.78 & 67.91
& 64.13 & 66.51
& 53.85 & 61.88 \\

Gemini-3.1 Flash & Detect
& \underline{84.13} & \underline{80.33}
& \underline{45.33} & \underline{55.73}
& 82.22 & 70.51
& 82.35 & 74.42
& 56.41 & 58.62
& 67.31 & 67.53 \\

\bottomrule
\end{tabular}
\end{adjustbox}

\caption{
Comparison of evaluation performance on the manually annotated evaluation set.
The table reports GUIJudge's process-aware performance, two ablations without trajectory retriever and without defect-specific verifiers, and comparisons with baseline evaluators that only support final-outcome detect evaluation.
Among GUIJudge and baseline evaluators (excluding ablations), bold marks the best result in each column, and \underline{underline} marks the second-best.
"--" indicates that the corresponding evaluator does not support that defect type.
}
\label{tab:rq1_guijudge_main}
\end{table*}

\section{Experiments}
We conduct comprehensive experiments to answer the following research questions:
\begin{itemize}[leftmargin=1em,itemsep=-2pt]
    \item \textbf{RQ1}: Does GUIJudge yield valid and reliable judgments of GUI testing outcomes?
    \item \textbf{RQ2}: How do existing systems perform in exploratory GUI testing, and what capability bottlenecks do they exhibit?
    \item \textbf{RQ3}: Can GUIJudge's defect verifiers improve the agent testing capability?
\end{itemize}
Accordingly, we first describe the experimental setup (\S 5.1), and then organize the experiments into three parts: GUIJudge validation (\S 5.2), model benchmarking on GUITestScape (\S 5.3), and verifier integration analysis (\S 5.4).


\subsection{Experiment Setup}
In this section, we describe experimental settings and evaluation metrics used in our experiments.

\noindent\textbf{RQ1: GUIJudge Validation.} To assess GUIJudge's design and judgment reliability, we construct a human-reference set. We compare GUIJudge against several representative evaluator baselines, including the hybrid evaluator from GUITester~\cite{gao2026guitester}, a rule-based evaluator, and three LLM-as-Judge models: GPT-5.4, Claude-4.6-opus, and Gemini-3.1-Flash. We further conduct two ablations on trajectory retrieval and defect-specific verifiers, together with backbone-sensitivity and evidence-removed analyses.

\noindent\textbf{RQ2: Benchmarking on GUITestScape.}
We benchmark various approaches on GUITestScape using GUIJudge as a unified evaluator, covering three categories: general-purpose models (General Models), including Claude-5.5-Opus, GPT-6-Astra, Gemini-3.1-Pro, Qwen-3-VL-Plus, Seed-1.8, and DeepSeek-4.1-Flash; 
GUI agents, including
UI-TARS-1.5-7B~\cite{qin2025uitarspioneeringautomatedgui},
GUI-Owl-1.5 (8B/32B)~\cite{xu2026mobileagentv35multiplatformfundamentalgui},
and MAI-UI-8B~\cite{zhou2025maiuitechnicalreportrealworld};
and coding agents, including Codex with GPT-6-Astra and Claude Code with Claude-5.5-opus and DeepSeek-4.1-flash backbones. This setup enables a systematic comparison of exploratory GUI testing capabilities across different fault modes and system categories.

\noindent\textbf{RQ3: Verifier Integration Analysis.} We integrate GUIJudge's defect-specific verifiers into UI-TARS, Seed-1.8, and Gemini-3.1-Pro to support defect recognition while retaining their navigation capabilities. We evaluate the resulting agents on GUITestScape through human evaluation.

\noindent\textbf{Evaluation Metrics.} Across all experiments, we adopt Recall and F1 as the primary metrics. Recall directly reflects a model's defect discovery capability in exploratory GUI testing, while F1 guards against overestimation arising from overly aggressive anomaly reporting.

\begin{table*}[t]
\centering
\small
\setlength{\tabcolsep}{4.0pt}
\renewcommand{\arraystretch}{1.12}

\begin{adjustbox}{max width=\textwidth}
\begin{tabular}{llcccccccccccc}
\toprule
\multirow{3}{*}{\textbf{System}} & \multirow{3}{*}{\textbf{Stage}}
& \multicolumn{4}{c}{\textbf{Display Defects}}
& \multicolumn{6}{c}{\textbf{Interaction Defects}}
& \multicolumn{2}{c}{\textbf{Overall}} \\
\cmidrule(lr){3-6} \cmidrule(lr){7-12} \cmidrule(lr){13-14}
&
& \multicolumn{2}{c}{\textbf{CR}}
& \multicolumn{2}{c}{\textbf{EL}}
& \multicolumn{2}{c}{\textbf{NLE}}
& \multicolumn{2}{c}{\textbf{ONR}}
& \multicolumn{2}{c}{\textbf{UTR}}
& \multicolumn{2}{c}{\textbf{Overall}} \\
\cmidrule(lr){3-4} \cmidrule(lr){5-6}
\cmidrule(lr){7-8} \cmidrule(lr){9-10}
\cmidrule(lr){11-12} \cmidrule(lr){13-14}
&
& Recall$\uparrow$ & F1$\uparrow$
& Recall$\uparrow$ & F1$\uparrow$
& Recall$\uparrow$ & F1$\uparrow$
& Recall$\uparrow$ & F1$\uparrow$
& Recall$\uparrow$ & F1$\uparrow$
& Recall$\uparrow$ & F1$\uparrow$ \\
\midrule

\rowcolor[gray]{0.94}
\multicolumn{14}{l}{\textbf{\textit{General Models}}} \\

Claude-5.5-Opus & Detect
& 51.09 & 63.09
& 43.60 & 56.82
& 54.10 & 64.71
& 49.33 & 57.81
& \underline{45.37} & \underline{57.65}
& 47.44 & 59.29 \\

GPT-6-Astra & Detect
& 56.41 & 65.67
& 46.91 & 59.38
& 46.15 & 59.26
& \underline{52.38} & 62.26
& 29.00 & 39.19
& 45.27 & 56.83 \\

Gemini-3.1-Pro & Detect
& 46.74 & 59.72
& 30.23 & 42.98
& 47.54 & 55.24
& 41.33 & 46.62
& 27.78 & 33.89
& 36.42 & 46.19 \\

Seed-1.8 & Detect
& 53.26 & 65.33
& 44.19 & 56.30
& 40.98 & 48.08
& 34.67 & 45.22
& 27.78 & 35.71
& 40.55 & 51.05 \\

Qwen-3-VL-Plus & Detect
& 28.26 & 42.62
& 22.09 & 33.63
& 27.87 & 37.36
& 28.00 & 33.87
& 14.81 & 20.51
& 23.23 & 32.82 \\

DeepSeek-4.1-Flash & Detect
& 44.57 & 57.75
& 32.56 & 43.92
& 11.48 & 20.29
& 20.27 & 32.61
& 11.43 & 19.67
& 25.99 & 38.53 \\

\midrule

\rowcolor[gray]{0.94}
\multicolumn{14}{l}{\textbf{\textit{GUI Agent Models}}} \\

UI-TARS-1.5-7B & Detect
& 5.43 & 9.71
& 1.74 & 3.28
& 29.51 & 37.89
& 8.00 & 10.34
& 5.56 & 7.74
& 7.48 & 11.66 \\

GUI-Owl-1.5-8B & Detect
& 1.09 & 2.13
& 1.74 & 3.39
& 1.64 & 2.99
& 0.00 & 0.00
& 0.93 & 1.75
& 1.18 & 2.26 \\

GUI-Owl-1.5-32B & Detect
& 9.78 & 17.65
& 3.49 & 6.63
& 4.92 & 8.70
& 2.67 & 4.82
& 0.00 & 0.00
& 3.94 & 7.30 \\

MAI-UI-8B & Detect
& 1.09 & 2.15
& 1.16 & 2.29
& 8.20 & 14.29
& 9.33 & 15.38
& 3.39 & 6.06
& 4.13 & 7.62 \\

\midrule

\rowcolor[gray]{0.94}
\multicolumn{14}{l}{\textbf{\textit{Coding Agents}}} \\

Codex (GPT-6-Astra) & Detect
& \textbf{65.22} & \textbf{74.53}
& \underline{50.58} & \underline{64.21}
& \underline{60.66} & \underline{70.48}
& \textbf{58.67} & \textbf{67.69}
& \underline{45.37} & 56.32
& \underline{54.53} & \underline{65.87} \\

Claude Code (Claude) & Detect
& \underline{62.09} & \underline{70.85}
& \textbf{53.62} & \textbf{69.81}
& \textbf{63.67} & \textbf{75.23}
& \underline{52.38} & 62.26
& \textbf{45.59} & \textbf{60.75}
& \textbf{57.02} & \textbf{69.44} \\

Claude Code (DeepSeek) & Detect
& 45.65 & 58.33
& 35.47 & 48.80
& 57.38 & 69.31
& 50.67 & \underline{63.33}
& 37.04 & 47.62
& 42.52 & 55.17 \\

\bottomrule
\end{tabular}
\end{adjustbox}

\caption{
Pass@1 Detect-stage performance of general models, GUI agent models, and coding agents on GUITestScape.
\textbf{Bold} marks the best result in each column, and \underline{underline} marks the second best.
}
\label{tab:rq2_model_performance}
\end{table*}

\subsection{RQ1}
\textbf{Overall Performance.} As shown in Table~\ref{tab:rq1_guijudge_main}, GUIJudge achieves 97.87\% and 97.12\% F1 scores at both the Reach and Trigger dimensions. At the Detect dimension, GUIJudge improves overall Recall from 73.21\% to 97.12\% (+23.91\%) and overall F1 from 81.19\% to 93.95\% (+12.76\%) over the strongest baseline evaluator. These results demonstrate that GUIJudge reliably retrieves defect-relevant trajectory segments and produces accurate judgments, enabling staged, process-aware evaluation beyond final-outcome assessment.

\noindent\textbf{Effect of Retrieval.} Removing trajectory retrieval reduces overall Detect Recall from 97.12\% to 76.92\%. The performance drop is smallest on ONR but largest on UTR. This is because ONR defects are typically triggered by a single action under relatively fixed verification conditions, making their evidence straightforward to localize; UTR defects, by contrast, often require cross-state comparison and delayed outcome verification, resulting in more dispersed and harder-to-localize evidence. These results indicate that evaluation reliability is fundamentally constrained by evidence localization quality, confirming that trajectory retrieval is essential for stable evaluation and constitutes a key source of GUIJudge's performance gains.

\noindent\textbf{Analysis Across Defect Types.} At the Detect stage, GUIJudge achieves the best performance across all defect types. By contrast, GUITester and the rule-based evaluator cannot handle display defects, while LLM-as-Judge methods exhibit unstable performance across defect categories. The ablation further reveals that replacing defect-specific verifiers with a unified verifier reduces Detect Recall from 97.12\% to 80.77\%, while also increasing performance variance across defect types. This degradation stems from the fundamentally different evidence requirements of the two defect categories: display defects are assessed through visual anomalies in interface states, whereas interaction defects are assessed through the consistency between action outcomes and task expectations. By tailoring each verifier to its corresponding evidence form, GUIJudge achieves more stable and reliable evaluation across diverse defect types.

\noindent\textbf{Additional Robustness Analyses.}
Additional analyses show that GUIJudge remains robust across evaluator backbones and does not produce positive \textit{Detect} judgments without supporting trajectory evidence (Appendix~\ref{sec:appendix:backbone_sensitivity}; Appendix~\ref{sec:appendix:shortcut}).


\vspace{0.1cm}
\begin{dialoguebox}{agentcolor}{ANSWER TO RQ1}
GUIJudge enables reliable process-aware evaluation through trajectory retrieval and defect-specific verification, with stable performance across evaluator backbones.
\end{dialoguebox}
\vspace{0.1cm}

\subsection{RQ2}
We benchmark 13 systems spanning general models, GUI agents, and coding agents on GUITestScape. Beyond overall defect-detection performance, we analyze stage-wise bottlenecks (Reach, Trigger, and Detect), fault-specific challenges, and multi-trial effects (Pass@1 vs. Pass@3).

\begin{figure*}[t]
    \centering
    \begin{subfigure}{0.24\textwidth}
        \centering
        \includegraphics[width=\linewidth]{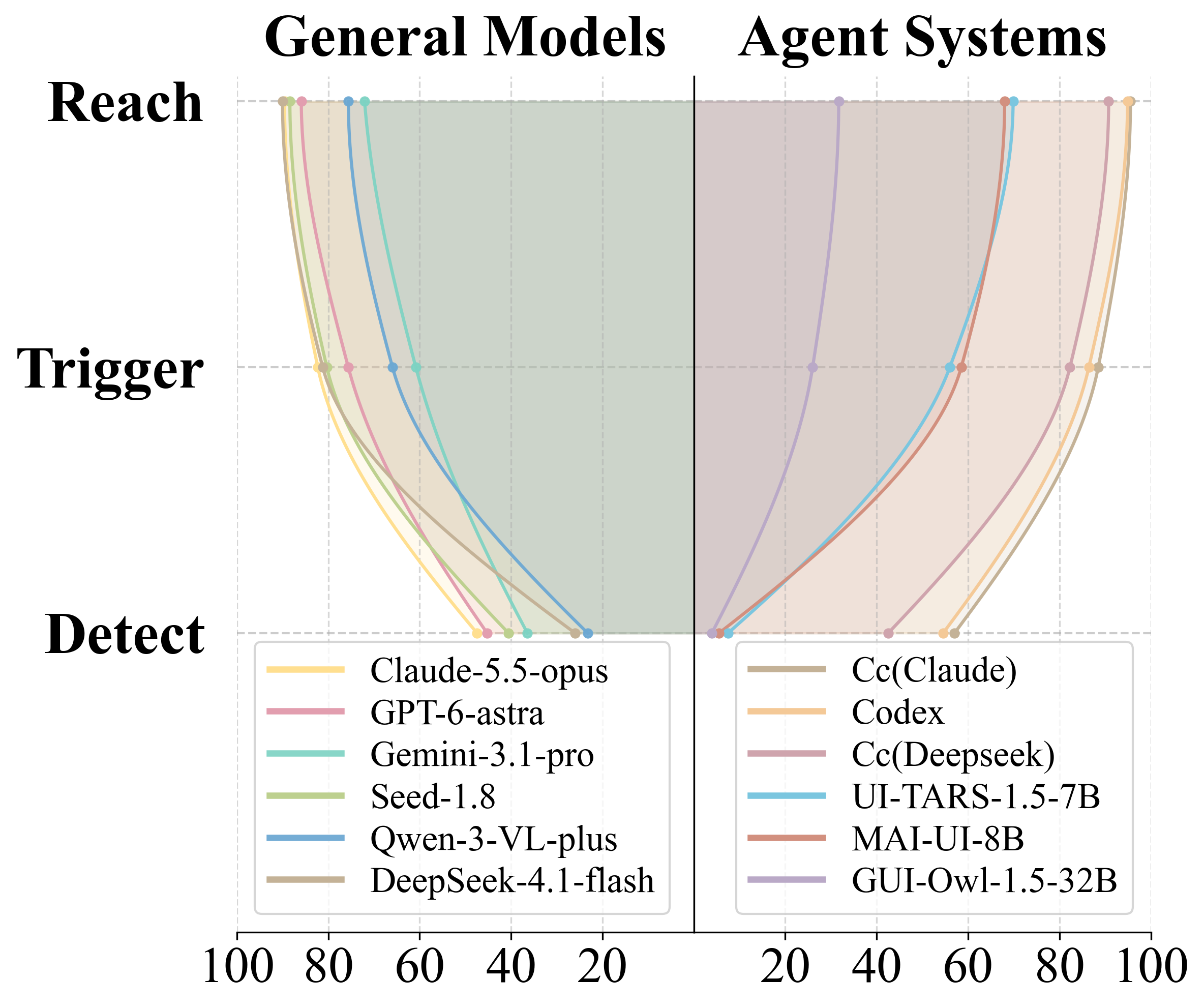}
        \caption{Overall Recall, Pass@1}
        \label{fig:rq2_recall_pass1}
    \end{subfigure}
    \hfill
    \begin{subfigure}{0.24\textwidth}
        \centering
        \includegraphics[width=\linewidth]{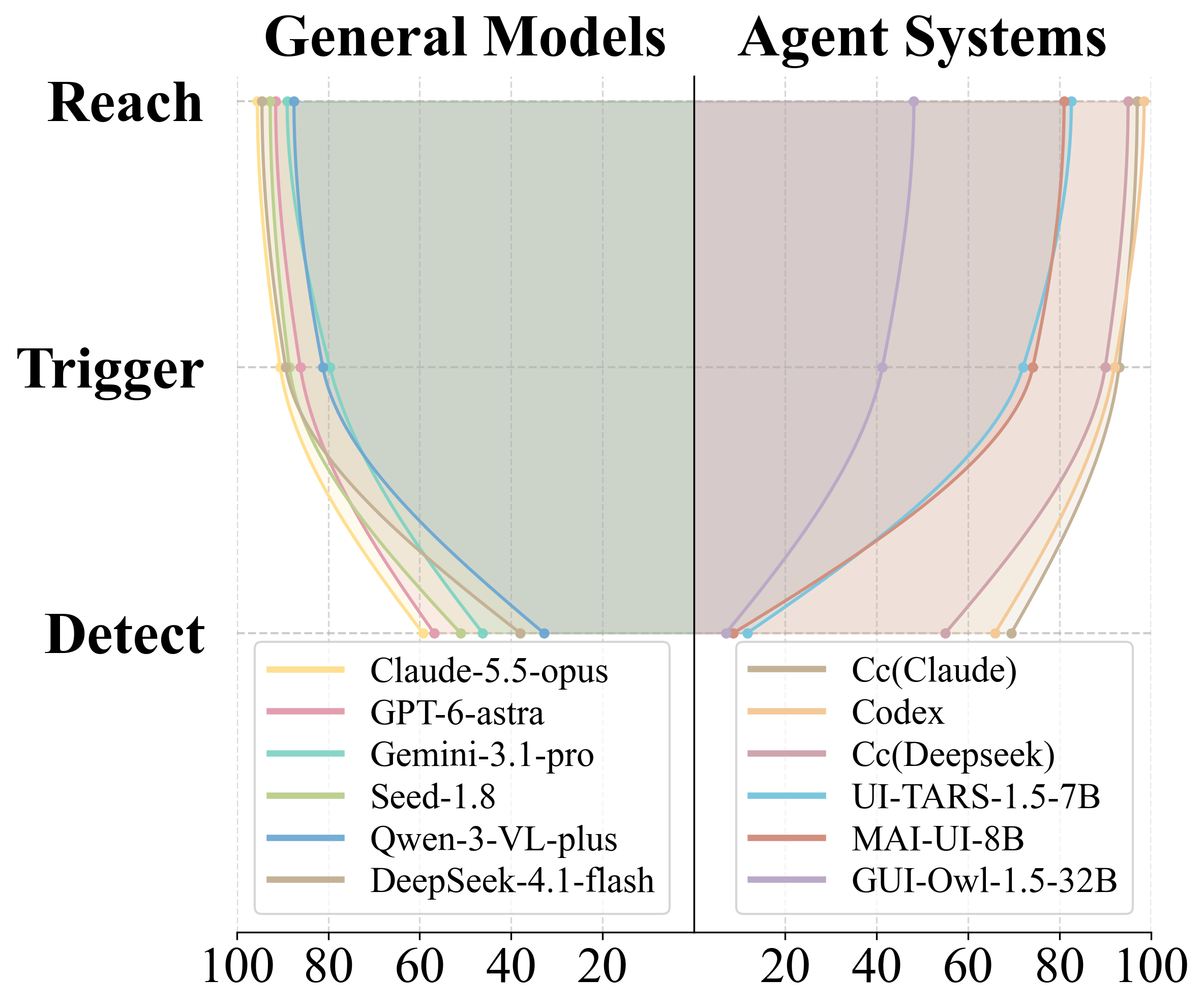}
        \caption{Overall F1, Pass@1}
        \label{fig:rq2_f1_pass1}
    \end{subfigure}
    \hfill
    \begin{subfigure}{0.24\textwidth}
        \centering
        \includegraphics[width=\linewidth]{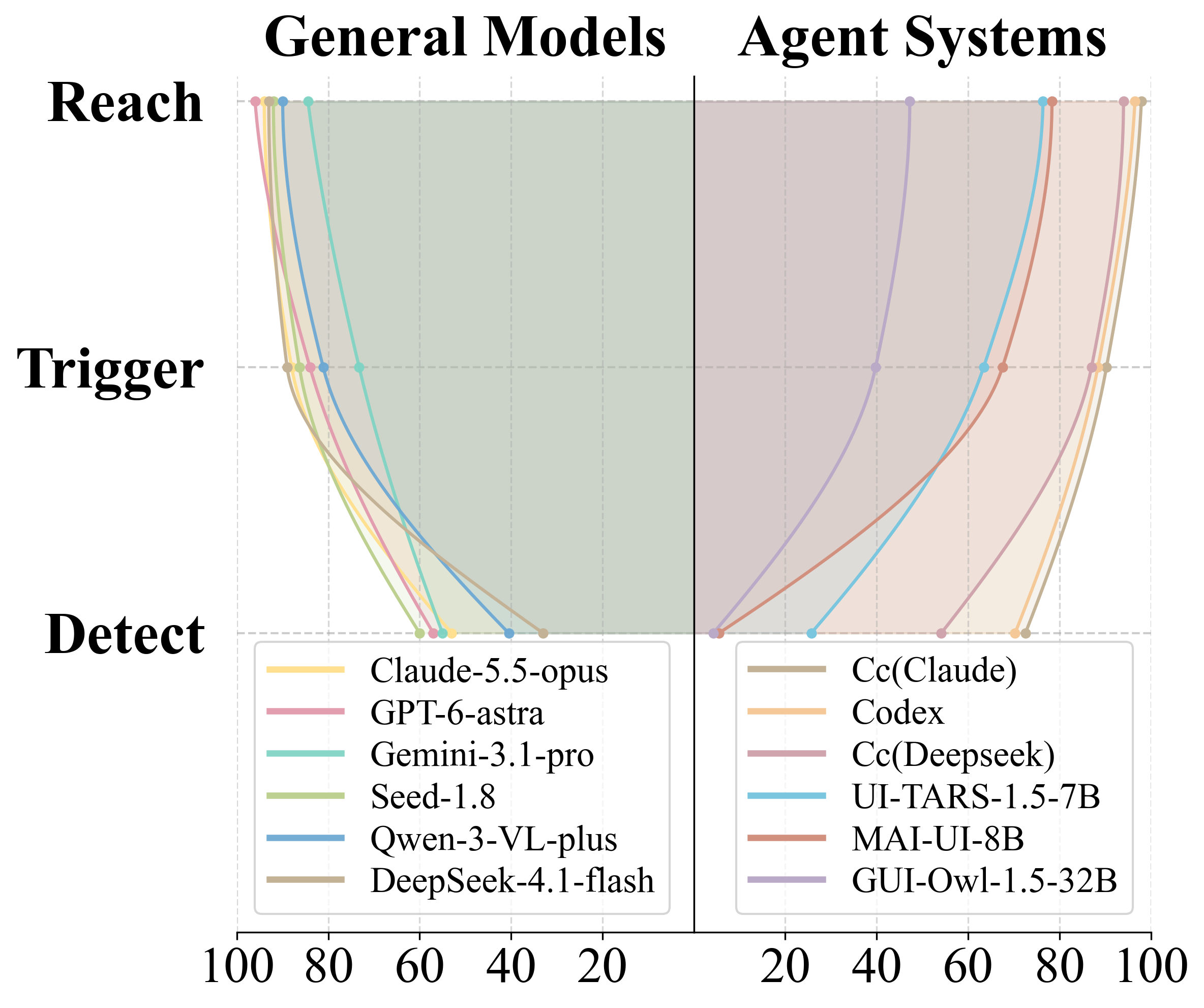}
        \caption{Overall Recall, Pass@3}
        \label{fig:rq2_recall_pass3}
    \end{subfigure}
    \hfill
    \begin{subfigure}{0.24\textwidth}
        \centering
        \includegraphics[width=\linewidth]{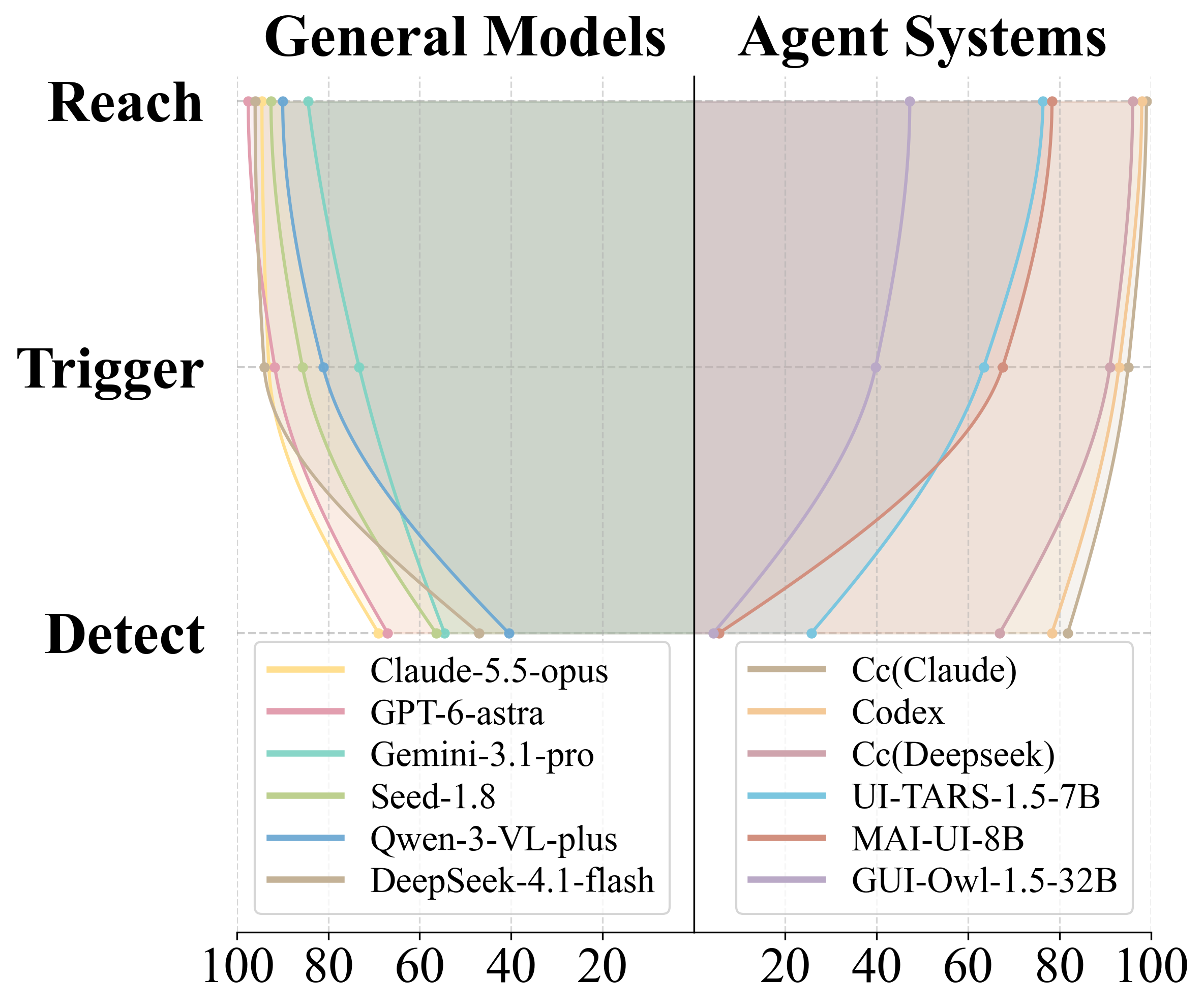}
        \caption{Overall F1, Pass@3}
        \label{fig:rq2_f1_pass3}
    \end{subfigure}

    \caption{
    Overall Recall and F1 trends across Reach, Trigger, and Detect for general models and agents on GUITestScape under Pass@1 and Pass@3. Detailed results are provided in Tables~\ref{tab:appendix_pass_at_1} and~\ref{tab:appendix_pass_at_3} in the appendix.
    }
    \label{fig:rq2_stagewise}
\end{figure*}




\noindent\textbf{Capability Bottlenecks Across System Categories.}
As shown in Figure~\ref{fig:rq2_stagewise} and Table~\ref{tab:appendix_pass_at_1}, under Pass@1, coding agents achieve the highest average Recall across Reach, Trigger, and Detect (93.72\%, 85.72\%, and 51.36\%), followed by general models (85.40\%, 75.50\%, and 36.48\%) and GUI agents (48.88\%, 40.27\%, and 4.18\%). Despite similar Trigger-to-Detect Recall drops, category-averaged Detect-to-Trigger Recall ratios differ substantially: 59.9\%, 48.3\%, and 10.4\%, respectively.

These results reveal distinct bottlenecks: GUI agents struggle with both navigation and defect detection, whereas general models and coding agents achieve stronger execution but remain limited in detection. Qualitative trajectory inspection further reveals a failure-attribution challenge: agents sometimes mistake genuine application defects for their own execution errors, leading to missed detections. Together, these findings highlight that successful navigation and defect triggering do not guarantee accurate defect recognition and reporting.

\noindent\textbf{Coding-Agent Gains and Backbone Dependence.}
As shown in Table~\ref{tab:appendix_pass_at_1}, under Pass@1, all three coding-agent configurations outperform their base models, with Detect F1 gains ranging from 9.04 to 16.64, compared with 0.59 to 4.94 at Reach and 1.17 to 6.60 at Trigger. Although coding agents achieve their largest gains at Detect, even the strongest configuration, Claude Code (Claude), reaches only 69.44\% Detect F1 versus 92.49\% at Trigger, revealing a persistent detection bottleneck.

Trajectory inspection explains these Detect gains: active observation helps avoid mistaking loading artifacts for display defects, iterative verification distinguishes execution errors from interaction defects, and context retention supports cross-state outcome verification. Yet some defects remain unrecognized despite evidence, suggesting that harnesses improve evidence acquisition, retention, and verification, while interpreting evidence, inferring expected behavior, and identifying defects still depend on the underlying model's reasoning and verification knowledge. Consistently, Table~\ref{tab:rq2_model_performance} shows that under the same framework, Claude achieves 69.44\% Detect F1 versus 55.17\% for DeepSeek, a 14.27-point gap. These findings suggest that coding-agent harnesses improve how GUI evidence is acquired and verified, but reliable defect recognition remains constrained by the underlying model's ability to interpret that evidence.

\noindent\textbf{Fault-Specific Detection Challenges.}
As shown in Table~\ref{tab:rq2_model_performance}, for display defects, 11 of 13 systems achieve higher Pass@1 Detect F1 on CR than EL. CR presents explicit visual anomalies, such as garbled text, whereas EL requires inferring plausible interface layouts that vary across applications, making structural plausibility harder to assess.

For interaction defects, all 13 systems achieve higher Detect F1 on NLE than UTR, and 12 of 13 perform better on ONR. NLE and ONR can often be judged from local action-state transitions, whereas UTR requires checking expected outcomes across states. These findings highlight two persistent limitations in current approaches: (1) judging interface structural plausibility and (2) verifying task outcomes under app-specific semantics.

\noindent\textbf{Multi-Trial Gains and Persistent Detection Bottlenecks.}
Figure~\ref{fig:rq2_stagewise} shows uneven Detect F1 gains from Pass@1 to Pass@3: 16.68 percentage points for general models on average, compared with 12.18 for coding agents and 6.56 for GUI agents. Tables~\ref{tab:appendix_pass_at_1} and~\ref{tab:appendix_pass_at_3} show that GUI-Owl-1.5-32B gains 15.67 points at Trigger but only 0.62 at Detect, indicating that improved defect triggering does not necessarily yield better detection.

As shown in Tables~\ref{tab:appendix_pass_at_1} and~\ref{tab:appendix_pass_at_3}, under the same framework, the Detect F1 gap between Claude and DeepSeek persists from Pass@1 (14.27 points) to Pass@3 (14.87 points), despite similar Pass@3 Trigger F1 (94.85\% vs. 93.39\%). These results suggest that repeated-trial gains remain constrained by the base model's defect-recognition capabilities.

\vspace{0.1cm}
\begin{dialoguebox}{agentcolor}{ANSWER TO RQ2}
(1) Existing systems exhibit substantial execution-to-detection gaps.
(2) Defect detection remains challenging in judging whether interface layouts are structurally plausible and verifying task outcomes under app-specific semantics.
(3) Coding-agent harnesses improve evidence acquisition, but reliable defect recognition remains constrained by the base model's ability to interpret GUI evidence, even with repeated trials.
\end{dialoguebox}
\vspace{0.1cm}


\subsection{RQ3}








\begin{table}[t]
\centering
\small
\setlength{\tabcolsep}{4pt}
\renewcommand{\arraystretch}{1.12}
\begin{tabular*}{\columnwidth}{@{\extracolsep{\fill}}lcc}
\toprule
\multirow{2}{*}{\textbf{Agent Setting}}
& \multicolumn{2}{c}{\textbf{Overall}} \\
\cmidrule(lr){2-3}
& \textbf{Recall$\uparrow$}
& \textbf{F1$\uparrow$} \\
\midrule
\rowcolor[gray]{0.94}
\multicolumn{3}{l}{\textit{w/ GUIJudge Verifier}} \\
UI-TARS-1.5-7B
& 57.02 \gain{+49.54}
& 69.07 \gain{+57.41} \\
Gemini-3.1-Pro
& \underline{61.27} \gain{+24.85}
& \underline{71.07} \gain{+24.88} \\
Seed-1.8
& \textbf{70.97} \gain{+30.42}
& \textbf{79.48} \gain{+28.43} \\
\bottomrule
\end{tabular*}
\caption{
Detect stage overall performance after integrating GUIJudge Verifier into each agent's workflow. Numbers in parentheses denote absolute gains over the corresponding unassisted baseline. \textbf{Bold} marks the best result, and \underline{underline} marks the second-best.
}
\label{tab:rq3_guijudge_assisted}
\end{table}

Table~\ref{tab:rq3_guijudge_assisted} shows that integrating GUIJudge's defect-specific verifiers substantially improves exploratory GUI testing performance across all evaluated agents. Most notably, UI-TARS-1.5-7B improves its Recall from 7.48\% to 57.02\% and its F1 from 11.66\% to 69.07\%. Since the verifiers are incorporated without retraining the underlying interaction policy, these gains indicate that a dedicated defect verification module can effectively compensate for agent models' inherent weakness in anomaly recognition and defect judgment after defect exposure.

\vspace{0.1cm}
\begin{dialoguebox}{agentcolor}{ANSWER TO RQ3}
Integrating GUIJudge's verifiers consistently improves Detect performance across all agents, confirming that defect recognition can be effectively decoupled from navigation capability and addressed through modular integration.
\end{dialoguebox}
\vspace{0.1cm}

\section{Conclusion}
This paper identifies two core challenges in exploratory GUI testing evaluation: insufficient coverage of display defects and closed-set final-outcome evaluation. We introduce GUITestScape, the first interactive benchmark jointly covering display and interaction defects, and propose GUIJudge, enabling open-set, process-aware evaluation. Experiments demonstrate the reliability of GUIJudge and further reveal that current GUI testing approaches mainly struggle with judging interface structural plausibility and verifying delayed outcomes.

\section*{Limitations}

(1) \textbf{Extraction of App-specific Test Basis.} 
Compared with closed-set evaluation, GUITestScape substantially reduces construction costs. However, reliable defect judgment requires app-specific standards about expected behavior and interface structure. In our benchmark, these standards are provided through resettable scenarios and scenario-level test bases, which support reproducible and judgeable evaluation, but their construction still partly relies on human effort. 

 \noindent(2) \textbf{Scope of Test Execution.}
GUITestScape focuses on the execution stage of exploratory GUI testing. The navigation tasks are assumed to be produced during the upstream test-design stage and provide high-level testing intents rather than fixed interaction paths. Given such tasks, GUITestScape evaluates whether an agent can reach, trigger, and detect defects through open-ended interaction. Test-task generation, boundary-condition design, and coverage planning therefore remain outside the current scope and constitute complementary upstream capabilities.

\noindent(3) \textbf{Platform Scope.}
The current benchmark and empirical evaluation are limited to mobile applications. However, GUI defects are common across mobile, web, and desktop interfaces rather than being specific to a single platform. Accordingly, the core evaluation design of this work is platform independent: the decomposition into Reach, Trigger, and Detect and GUIJudge's open-set evaluation are applicable to other platforms. Platform dependence is confined to the underlying execution layer, such as environment control, GUI observation, and action execution. Empirical validation on web and desktop environments remains future work.


\noindent(4) \textbf{From Evaluation to Model Improvement.}
Our RQ2 results reveal persistent execution-to-detection gaps, even with coding-agent harnesses and repeated trials. While RQ3 demonstrates that integrating GUIJudge's verifiers improves system-level detection performance without retraining, this modular enhancement does not directly improve the underlying models' intrinsic defect-recognition capabilities. Developing training environments with scalable interaction tasks and process-level rewards to strengthen these capabilities remains an important direction for future work.

Future work will focus on the following four research directions:
(1) developing training environments with scalable interaction tasks and process-level rewards to improve agents' defect-recognition capabilities,
(2) automating test basis construction from project artifacts and interaction evidence, 
(3) extending evaluation from test execution to test design, including task generation and boundary condition design, and 
(4) validating the proposed evaluation framework on web and desktop environments and adapting the underlying execution infrastructure accordingly.




\bibliography{custom}

\appendix
\section{Test-Basis Construction and Analysis}

\subsection{Test-Basis Construction}
The test basis is grounded in developer-authored artifacts rather than written specifically for GUIJudge. The preset defects in GUITestScape are collected from public issue trackers and bug reports, where developers describe the expected behavior, observed behavior, reproduction context, and relevant application logic. We extract the underlying design specifications and business logic from these artifacts and organize them into scenario-level correctness criteria. Each preset defect therefore has at least one applicable criterion grounded in the original developer feedback, while the criterion itself does not encode the defect instance or prescribe its manifestation. Rather than prescribing a unique action trajectory, the test basis provides app-specific semantics while remaining permissive enough to accommodate diverse interaction patterns.

\subsection{Test-Basis Shortcut Controls}
\label{sec:appendix:shortcut}
To examine whether GUIJudge can exploit the scenario-level test basis as a shortcut, we construct an evidence-removed negative control.
We retain the navigation task, scenario-level test basis, and original agent defect report, while removing the defect-relevant trajectory evidence.
Under this setting, the number of positive \textit{Detect} judgments drops to zero.
This result indicates that the test basis alone does not function as a defect answer and that GUIJudge requires supporting evidence from the observed trajectory to produce a positive judgment.

\section{GUIJudge}
\label{sec:appendix}

\subsection{Workflow of GUIJudge}
\label{sec:appendix:workflow}

Algorithm~\ref{alg:guijudge_workflow} presents the workflow of GUIJudge for evaluating a single agent trajectory.
GUIJudge takes as input the agent trajectory $\tau$, the agent defect report $\rho$, the navigation task $g$, and the corresponding scenario-level test basis $B$.
The test basis $B$ contains a set of app-specific correctness criteria, each specifying a precondition, a trigger condition, and expected evidence.

GUIJudge proceeds in two stages.
First, the \textbf{Trajectory Retriever} selects from $B$ the correctness criterion $c^{*}$ that is applicable to the current task and observed trajectory, and then retrieves defect-relevant segments according to the criterion's semantic conditions.
Second, the retrieved segments are passed to two defect-specific verifiers:
the \textbf{Display Defect Verifier} examines GUI states for display defects, while the \textbf{Interaction Defect Verifier} examines action--state transitions for interaction defects.
The workflow finally returns the verdicts \textit{Reach}, \textit{Trigger}, and \textit{Detect} for the trajectory.

\noindent\textbf{Trajectory Retriever}
first selects the applicable correctness criterion
$c^{*} \in B$
based on the navigation task $g$ and the observed trajectory $\tau$.
The selected criterion provides three semantic conditions:
$c^{*}_{\text{precondition}}$,
$c^{*}_{\text{trigger}}$, and
$c^{*}_{\text{evidence}}$.
The retriever then scans the full trajectory to localize candidate segments that instantiate these conditions.
Once a trajectory step matches
$c^{*}_{\text{precondition}}$,
GUIJudge marks \textit{Reach} as true.
It subsequently searches forward for the interaction or event corresponding to
$c^{*}_{\text{trigger}}$
and the resulting state that matches
$c^{*}_{\text{evidence}}$.
When these conditions are grounded in temporal order, the corresponding sub-trajectory is recorded as a defect-relevant segment.
The retrieved segments are treated only as candidate evidence and are passed to the defect-specific verifiers for final judgment.

\noindent\textbf{Display Defect Verifier}
treats \textit{Trigger} as not applicable and sets it to $\bot$.
If the retrieved defect-relevant segments are non-empty, the Display Defect Verifier examines each state within the retrieved segments under the selected correctness criterion $c^{*}$.
Once a state is verified to violate the expected GUI behavior specified by $c^{*}$, GUIJudge compares the verified defect with the agent defect report $\rho$.
If the reported step and defect description are consistent with the verified result, \textit{Detect} is marked as true.

\noindent\textbf{Interaction Defect Verifier}
verifies the retrieved defect-relevant segments at the level of state transitions under the selected correctness criterion $c^{*}$.
For each segment, it resets the interaction history $H$ and traverses all transitions in temporal order.
After each transition is added to the history, the verifier judges whether the observed response violates the behavior specified by $c^{*}$, given the navigation task $g$ and the accumulated interaction history $H$.
Once an interaction defect is verified, \textit{Trigger} is marked as true.
GUIJudge then compares the verified defect with the agent defect report $\rho$.
If they are consistent, \textit{Detect} is marked as true.

Together, the Trajectory Retriever and the two defect-specific verifiers produce the final \textit{Reach}, \textit{Trigger}, and \textit{Detect} verdicts for the trajectory, which are then used by the downstream metrics in Section~5.

\begin{algorithm}[!ht]
\begin{minipage}{0.95\linewidth}
    \footnotesize
    \setlength{\algomargin}{0.5em}
    \caption{GUIJudge Workflow.}
    \SetAlgoVlined
    \SetInd{0.3em}{0.6em}
    \label{alg:guijudge_workflow}

    \KwIn{
    agent trajectory $\tau$,
    agent defect report $\rho$,
    navigation task $g$,
    scenario-level test basis
    $B=\{c_1,\ldots,c_m\}$
    }
    \KwOut{verdicts $(\text{reach},\,\text{trigger},\,\text{detect})$}

    \LinesNotNumbered
    \SetKwInput{KwInit}{Init}
    \KwInit{
    $\text{reach} \leftarrow \texttt{False}$;
    $\text{trigger} \leftarrow \texttt{False}$; \\
    $\text{detect} \leftarrow \texttt{False}$;
    $\text{segments} \leftarrow \emptyset$.
    }
    \LinesNumbered

    \BlankLine
    \tcp{\textcolor{blue}{Stage 1a: Correctness-Criterion Selection}}
    $c^{*} \leftarrow \textsc{SelectCriterion}(g,\,\tau,\,B)$\;

    \BlankLine
    \tcp{\textcolor{blue}{Stage 1b: Trajectory Retrieval}}
    $t \leftarrow 1$\;
    \While{$t \le |\tau|$}{
        \If{$\textsc{Match}(\tau[t],\,c^{*}_{\text{precondition}})$}{
            $\text{reach} \leftarrow \texttt{True}$\;
            $s \leftarrow t$\;
            $u \leftarrow t + 1$\;

            \While{$u \le |\tau|$}{
                \If{$\textsc{Match}(\tau[u],\,c^{*}_{\text{trigger}})$}{
                    $v \leftarrow u$\;

                    \While{$v \le |\tau|$}{
                        \If{$\textsc{Match}(\tau[v],\,c^{*}_{\text{evidence}})$}{
                            $\text{segments}
                            \leftarrow
                            \text{segments} \cup \{\tau[s:v]\}$\;
                            $t \leftarrow v$\;
                            \textbf{break}\;
                        }
                        $v \leftarrow v + 1$\;
                    }
                }

                \If{$t = v$}{
                    \textbf{break}\;
                }

                $u \leftarrow u + 1$\;
            }
        }
        $t \leftarrow t + 1$\;
    }

    \BlankLine
    \tcp{\textcolor{blue}{Stage 2a: Display Defect Verifier}}
    \If{$\text{segments} \neq \emptyset$}{
        \ForEach{segment $\sigma \in \text{segments}$}{
            \ForEach{state $o \in \sigma$}{
                $r \leftarrow
                f_{\theta}^{\text{disp}}(o,\,c^{*})$\;

                \If{$r.\texttt{has\_defect}$}{
                    $\text{trigger}
                    \leftarrow
                    \bot$
                    \tcp*[r]{\textcolor{gray}{not applicable for display defects}}

                    \If{$\textsc{Consistent}(\rho, r)$}{
                        $\text{detect}
                        \leftarrow
                        \texttt{True}$\;
                    }
                }
            }
        }
    }

    \BlankLine
    \tcp{\textcolor{blue}{Stage 2b: Interaction Defect Verifier}}
    \If{$\text{segments} \neq \emptyset$}{
        \ForEach{segment $\sigma \in \text{segments}$}{
            $H \leftarrow \emptyset$\;
            $\Delta \leftarrow \textsc{Transitions}(\sigma)$\;

            \ForEach{transition $(o_t,a_t,o_{t+1}) \in \Delta$}{
                $H \leftarrow
                H \cup \{(o_t,a_t,o_{t+1})\}$\;

                $r_t \leftarrow
                f_{\theta}^{\text{int}}
                (g,\,c^{*},\,H,\,(o_t,a_t,o_{t+1}))$\;

                \If{$r_t.\texttt{has\_defect}$}{
                    $\text{trigger}
                    \leftarrow
                    \texttt{True}$\;

                    \If{$\textsc{Consistent}(\rho,r_t)$}{
                        $\text{detect}
                        \leftarrow
                        \texttt{True}$\;
                    }
                }
            }
        }
    }

    \Return $(\text{reach},\,\text{trigger},\,\text{detect})$\;
\end{minipage}
\end{algorithm}

  \subsection{Prompt}
  \label{sec:appendix:prompt}

  Each module of GUIJudge is driven by a dedicated prompt. The prompt used by the trajectory retriever is shown in Table~\ref{prompt:retrieval}, the prompt used by the display verifier is shown in Table~\ref{prompt:display}, and the prompt used by the interaction verifier is shown in Table~\ref{prompt:interaction}.

\subsection{GUIJudge Backbone Sensitivity}
\label{sec:appendix:backbone_sensitivity}
To examine whether GUIJudge depends on a particular backbone model, we rerun the complete RQ1 evaluation while keeping the evaluation pipeline, prompts, and data unchanged and replacing only the backbone model. In addition to the default Claude-4.6-Opus, we evaluate Gemini-3.1-Pro, GPT-5.4, and the local Qwen-3.6-27B model. As shown in Table~\ref{tab:backbone_sensitivity}, GUIJudge maintains strong Detect performance across all four backbones, while Claude-4.6-Opus achieves the best overall result and is therefore used as the default backbone.

We further measure pairwise consistency of the Detect judgments produced by different backbones. Table~\ref{tab:backbone_consistency} shows raw agreement of 91.09\%--95.54\% and Cohen's $\kappa$ of 0.822--0.911. These results indicate that GUIJudge's judgments are stable across substantially different backbone models rather than being specific to Claude-4.6-Opus.

\begin{table*}[t]
\centering
\small
\setlength{\tabcolsep}{4.5pt}
\renewcommand{\arraystretch}{1.12}

\begin{adjustbox}{max width=\textwidth}
\begin{tabular}{lcccccc}
\toprule
\textbf{Backbone} &
\textbf{CR R/F1 (\%)} &
\textbf{EL R/F1 (\%)} &
\textbf{NLE R/F1 (\%)} &
\textbf{ONR R/F1 (\%)} &
\textbf{UTR R/F1 (\%)} &
\textbf{Overall R/F1 (\%)} \\
\midrule

Claude-4.6-Opus (default)
& 100.00/89.36
& 96.00/92.31
& 100.00/96.77
& 100.00/100.00
& 92.31/94.12
& \textbf{97.12/93.95} \\

Gemini-3.1-Pro
& 85.71/83.72
& 84.00/87.50
& 100.00/96.77
& 100.00/100.00
& 96.15/98.04
& 92.31/92.75 \\

GPT-5.4
& 90.48/86.36
& 80.00/85.11
& 93.33/93.33
& 94.12/94.12
& 96.15/98.04
& 90.38/91.26 \\

Qwen-3.6-27B (local)
& 85.71/80.00
& 76.00/82.61
& 93.33/93.33
& 100.00/100.00
& 88.46/92.00
& 87.50/88.78 \\

\bottomrule
\end{tabular}
\end{adjustbox}

\caption{
Backbone sensitivity of GUIJudge on the RQ1 human-reference set.
We report Detect Recall/F1 (\%) for each fault mode and overall performance.
}
\label{tab:backbone_sensitivity}
\end{table*}

\begin{table}[t]
\centering
\footnotesize
\setlength{\tabcolsep}{3pt}
\renewcommand{\arraystretch}{1.12}

\begin{tabular}{lcccc}
\toprule
& \textbf{Claude}
& \textbf{Gemini}
& \textbf{GPT-5.4}
& \textbf{Qwen} \\
\midrule

Claude-4.6-Opus
& --
& 0.862
& 0.832
& 0.822 \\

Gemini-3.1-Pro
& 93.07
& --
& 0.911
& 0.881 \\

GPT-5.4
& 91.58
& 95.54
& --
& 0.871 \\

Qwen-3.6-27B
& 91.09
& 94.06
& 93.56
& -- \\

\bottomrule
\end{tabular}

\caption{
Cross-backbone consistency of GUIJudge on Detect judgments.
The lower triangle reports pairwise raw agreement (\%), and the upper triangle reports Cohen's $\kappa$.
}
\label{tab:backbone_consistency}
\end{table}

\section{Additional Experimental Details and Results}
\label{sec:appendix:additional_results}

\subsection{Human-Reference Set Construction}
\label{app:human_reference}

The human-reference set used in RQ1 is constructed from actual agent executions on GUITestScape rather than from GUIJudge outputs. Following the evaluation cases defined in \S3.2, we run GUI agents on the navigation tasks under the corresponding resettable scenarios and collect their complete execution trajectories.

Two annotators with Android GUI testing experience independently inspect each trajectory and assign \textit{Reach}, \textit{Trigger}, and \textit{Detect} labels according to the definitions in \S4.2. The task and test basis provide the testing intent and app-specific correctness criteria, while the labels are determined from the observed trajectory and agent report. The annotators do not access GUIJudge predictions during annotation.

We compute inter-annotator agreement before adjudication. As shown in Table~\ref{tab:iaa}, raw agreement ranges from 95.05\% to 98.51\%, with Cohen's $\kappa$ ranging from 0.9010 to 0.9545. Disagreements are subsequently resolved through discussion to obtain the final reference labels used in RQ1.

\begin{table}[t]
\centering
\small
\setlength{\tabcolsep}{10pt}
\renewcommand{\arraystretch}{1.15}

\begin{tabular}{lcc}
\toprule
\textbf{Stage} & \textbf{Raw Agreement (\%)} & \textbf{Cohen's $\kappa$} \\
\midrule
Reach   & 98.51 & 0.9545 \\
Trigger & 97.03 & 0.9391 \\
Detect  & 95.05 & 0.9010 \\
\bottomrule
\end{tabular}

\caption{Inter-annotator agreement on the RQ1 human-reference set.}
\label{tab:iaa}
\end{table}

  \subsection{Information Access}
  \label{sec:appendix:information_access}

GUIJudge plays different roles across RQ1--RQ3. Table~\ref{tab:guijudge_access} summarizes its role and available information during execution and scoring.

\begin{table}[t]
\centering
\small
\setlength{\tabcolsep}{3pt}
\renewcommand{\arraystretch}{1.15}

\begin{tabular}{
    @{}l
    >{\raggedright\arraybackslash}p{2.45cm}
    >{\raggedright\arraybackslash}p{4.55cm}
    @{}
}
\toprule
\textbf{RQ} &
\textbf{Usage Phase} &
\textbf{Available Information} \\
\midrule

RQ1 &
Scoring 
{\footnotesize \gain{Evaluator under validation}} &
Full trajectory, agent report, task, and scenario-level test basis \\

RQ2 &
Scoring \newline
{\footnotesize \gain{Unified evaluator}} &
Full trajectory, agent report, task, and scenario-level test basis \\

RQ3 &
Execution \newline
{\footnotesize \gain{Integrated verifier}} &
Task instruction, GUI observation, and action history \\

\bottomrule
\end{tabular}

\caption{
Roles and information access of GUIJudge across RQ1--RQ3.
}
\label{tab:guijudge_access}
\end{table}

  \subsection{Process-Level Metric: Pass@1}
  \label{sec:appendix:pass_at_1}

Table~\ref{tab:appendix_pass_at_1} reports Pass@1 results on GUITestScape across the Reach, Trigger, and Detect stages and five fault modes for both general models and specialized agent models.

\begin{table*}[t]
\centering
\small
\setlength{\tabcolsep}{4.0pt}
\renewcommand{\arraystretch}{1.12}

\begin{adjustbox}{max width=\textwidth}
\begin{tabular}{llcccccccccccc}
\toprule
\multirow{3}{*}{\textbf{Model}} & \multirow{3}{*}{\textbf{Type}}
& \multicolumn{4}{c}{\textbf{Display Defects}}
& \multicolumn{6}{c}{\textbf{Interaction Defects}}
& \multicolumn{2}{c}{\textbf{Overall}} \\
\cmidrule(lr){3-6} \cmidrule(lr){7-12} \cmidrule(lr){13-14}
&
& \multicolumn{2}{c}{\textbf{CR}}
& \multicolumn{2}{c}{\textbf{EL}}
& \multicolumn{2}{c}{\textbf{NLE}}
& \multicolumn{2}{c}{\textbf{ONR}}
& \multicolumn{2}{c}{\textbf{UTR}}
& \multicolumn{2}{c}{\textbf{Overall}} \\
\cmidrule(lr){3-4} \cmidrule(lr){5-6}
\cmidrule(lr){7-8} \cmidrule(lr){9-10}
\cmidrule(lr){11-12} \cmidrule(lr){13-14}
&
& Recall$\uparrow$ & F1$\uparrow$
& Recall$\uparrow$ & F1$\uparrow$
& Recall$\uparrow$ & F1$\uparrow$
& Recall$\uparrow$ & F1$\uparrow$
& Recall$\uparrow$ & F1$\uparrow$
& Recall$\uparrow$ & F1$\uparrow$ \\
\midrule

\rowcolor[gray]{0.94}
\multicolumn{14}{l}{\textit{\textbf{Reach Stage}}} \\

Claude-5.5-Opus & General
& 91.30 & 95.45
& 92.44 & 96.07
& 85.25 & 92.04
& 78.67 & 88.06
& \textbf{93.52} & \textbf{96.65}
& 89.57 & 94.50 \\

GPT-6-Astra & General
& 85.90 & 92.41
& 89.51 & 94.46
& 78.85 & 88.17
& 80.95 & 89.47
& 87.00 & 93.05
& 85.93 & 92.43 \\

Gemini-3.1-Pro & General
& 80.43 & 89.16
& 83.14 & 90.79
& 77.05 & 87.04
& 78.67 & 88.06
& 75.93 & 86.32
& 79.72 & 88.72 \\

Seed-1.8 & General
& 89.13 & 94.25
& 90.12 & 94.80
& 85.25 & 92.04
& 86.67 & 92.86
& 87.96 & 93.60
& 88.39 & 93.83 \\

Qwen-3-VL-Plus & General
& 73.91 & 85.00
& 84.88 & 91.82
& 80.33 & 89.09
& 73.33 & 84.62
& 77.78 & 87.50
& 79.13 & 88.35 \\

DeepSeek-4.1-Flash & General
& 92.39 & 96.05
& 93.60 & 96.70
& 85.25 & 92.04
& 78.38 & 87.88
& 91.43 & 95.52
& 89.68 & 94.56 \\

UI-TARS-1.5-7B & GUI Agent
& 52.17 & 68.57
& 76.74 & 86.84
& 67.21 & 80.39
& 61.33 & 76.03
& 73.15 & 84.49
& 68.11 & 81.03 \\

GUI-Owl-1.5-8B & GUI Agent
& 19.78 & 33.03
& 31.98 & 48.46
& 21.31 & 35.14
& 21.33 & 35.16
& 25.93 & 41.18
& 25.64 & 40.82 \\

GUI-Owl-1.5-32B & GUI Agent
& 40.22 & 57.36
& 37.21 & 54.24
& 32.79 & 49.38
& 21.33 & 35.16
& 22.22 & 36.36
& 31.69 & 48.13 \\

MAI-UI-8B & GUI Agent
& 66.30 & 79.74
& 66.86 & 80.14
& 77.05 & 87.04
& 62.67 & 77.05
& 79.63 & 88.66
& 70.08 & 82.41 \\

Codex (GPT-6-Astra) & Coding Agent
& \textbf{98.91} & \textbf{99.45}
& \underline{95.35} & \underline{97.62}
& \underline{96.72} & \underline{98.33}
& \underline{89.33} & \textbf{94.37}
& \textbf{93.52} & \textbf{96.65}
& \underline{94.88} & \textbf{97.37} \\

Claude Code (Claude) & Coding Agent
& \underline{98.37} & \underline{99.12}
& \textbf{97.09} & \textbf{98.53}
& \textbf{97.54} & \textbf{98.75}
& \textbf{91.35} & \underline{93.46}
& 92.41 & 94.97
& \textbf{95.52} & \underline{97.17} \\

Claude Code (DeepSeek) & Coding Agent
& 91.30 & 95.45
& 94.77 & 97.31
& 85.25 & 92.04
& 82.67 & 90.51
& \underline{92.59} & \underline{96.15}
& 90.75 & 95.15 \\

\midrule

\rowcolor[gray]{0.94}
\multicolumn{14}{l}{\textit{\textbf{Trigger Stage}}} \\

Claude-5.5-Opus & General
& 91.30 & 95.45
& 92.44 & 96.07
& 72.13 & 83.81
& 68.00 & 80.95
& \underline{74.07} & \textbf{85.11}
& 82.28 & 90.28 \\

GPT-6-Astra & General
& 85.90 & 92.41
& 89.51 & 94.46
& 61.54 & 76.19
& 71.43 & 83.33
& 55.00 & 70.97
& 75.60 & 86.11 \\

Gemini-3.1-Pro & General
& 80.43 & 89.16
& 83.14 & 90.79
& 57.38 & 72.92
& 54.67 & 70.69
& 39.81 & 56.95
& 66.14 & 79.62 \\

Seed-1.8 & General
& 89.13 & 94.25
& 90.12 & 94.80
& 80.33 & \underline{89.09}
& 69.33 & 81.89
& 64.81 & 78.65
& 80.31 & 89.08 \\

Qwen-3-VL-Plus & General
& 73.91 & 85.00
& 84.88 & 91.82
& 63.93 & 78.00
& 56.00 & 71.79
& 48.15 & 65.00
& 68.31 & 81.17 \\

DeepSeek-4.1-Flash & General
& 92.39 & 96.05
& 93.60 & 96.70
& 68.85 & 81.55
& 64.86 & 78.69
& 65.71 & 79.31
& 80.36 & 89.11 \\

UI-TARS-1.5-7B & GUI Agent
& 52.17 & 68.57
& 76.74 & 86.84
& 45.90 & 62.92
& 30.67 & 46.94
& 44.44 & 61.54
& 54.92 & 70.45 \\

GUI-Owl-1.5-8B & GUI Agent
& 19.78 & 33.03
& 29.65 & 45.74
& 21.31 & 35.14
& 16.00 & 27.59
& 10.19 & 18.49
& 20.71 & 34.31 \\

GUI-Owl-1.5-32B & GUI Agent
& 40.22 & 57.36
& 37.21 & 54.24
& 29.51 & 45.57
& 10.67 & 19.28
& 4.63 & 8.85
& 25.98 & 41.25 \\

MAI-UI-8B & GUI Agent
& 66.30 & 79.74
& 66.86 & 80.14
& 60.66 & 75.51
& 48.00 & 64.86
& 49.07 & 65.84
& 59.45 & 74.57 \\

Codex (GPT-6-Astra) & Coding Agent
& \textbf{98.91} & \textbf{99.45}
& \underline{95.35} & \underline{97.62}
& \textbf{81.97} & \textbf{90.09}
& \underline{78.67} & \underline{88.06}
& 69.44 & 81.97
& \underline{86.42} & \textbf{92.71} \\

Claude Code (Claude) & Coding Agent
& \underline{98.37} & \underline{99.12}
& \textbf{97.09} & \textbf{98.53}
& \underline{80.98} & 85.77
& \textbf{82.41} & \textbf{90.36}
& \textbf{74.73} & \underline{83.62}
& \textbf{88.47} & \underline{92.49} \\

Claude Code (DeepSeek) & Coding Agent
& 91.30 & 95.45
& 94.77 & 97.31
& 75.41 & 85.98
& 70.67 & 82.81
& 66.67 & 80.00
& 82.28 & 90.28 \\

\midrule

\rowcolor[gray]{0.94}
\multicolumn{14}{l}{\textit{\textbf{Detect Stage}}} \\

Claude-5.5-Opus & General
& 51.09 & 63.09
& 43.60 & 56.82
& 54.10 & 64.71
& 49.33 & 57.81
& \underline{45.37} & \underline{57.65}
& 47.44 & 59.29 \\

GPT-6-Astra & General
& 56.41 & 65.67
& 46.91 & 59.38
& 46.15 & 59.26
& \underline{52.38} & 62.26
& 29.00 & 39.19
& 45.27 & 56.83 \\

Gemini-3.1-Pro & General
& 46.74 & 59.72
& 30.23 & 42.98
& 47.54 & 55.24
& 41.33 & 46.62
& 27.78 & 33.89
& 36.42 & 46.19 \\

Seed-1.8 & General
& 53.26 & 65.33
& 44.19 & 56.30
& 40.98 & 48.08
& 34.67 & 45.22
& 27.78 & 35.71
& 40.55 & 51.05 \\

Qwen-3-VL-Plus & General
& 28.26 & 42.62
& 22.09 & 33.63
& 27.87 & 37.36
& 28.00 & 33.87
& 14.81 & 20.51
& 23.23 & 32.82 \\

DeepSeek-4.1-Flash & General
& 44.57 & 57.75
& 32.56 & 43.92
& 11.48 & 20.29
& 20.27 & 32.61
& 11.43 & 19.67
& 25.99 & 38.53 \\

UI-TARS-1.5-7B & GUI Agent
& 5.43 & 9.71
& 1.74 & 3.28
& 29.51 & 37.89
& 8.00 & 10.34
& 5.56 & 7.74
& 7.48 & 11.66 \\

GUI-Owl-1.5-8B & GUI Agent
& 1.09 & 2.13
& 1.74 & 3.39
& 1.64 & 2.99
& 0.00 & 0.00
& 0.93 & 1.75
& 1.18 & 2.26 \\

GUI-Owl-1.5-32B & GUI Agent
& 9.78 & 17.65
& 3.49 & 6.63
& 4.92 & 8.70
& 2.67 & 4.82
& 0.00 & 0.00
& 3.94 & 7.30 \\

MAI-UI-8B & GUI Agent
& 1.09 & 2.15
& 1.16 & 2.29
& 8.20 & 14.29
& 9.33 & 15.38
& 3.39 & 6.06
& 4.13 & 7.62 \\

Codex (GPT-6-Astra) & Coding Agent
& \textbf{65.22} & \textbf{74.53}
& \underline{50.58} & \underline{64.21}
& \underline{60.66} & \underline{70.48}
& \textbf{58.67} & \textbf{67.69}
& \underline{45.37} & 56.32
& \underline{54.53} & \underline{65.87} \\

Claude Code (Claude) & Coding Agent
& \underline{62.09} & \underline{70.85}
& \textbf{53.62} & \textbf{69.81}
& \textbf{63.67} & \textbf{75.23}
& \underline{52.38} & 62.26
& \textbf{45.59} & \textbf{60.75}
& \textbf{57.02} & \textbf{69.44} \\

Claude Code (DeepSeek) & Coding Agent
& 45.65 & 58.33
& 35.47 & 48.80
& 57.38 & 69.31
& 50.67 & \underline{63.33}
& 37.04 & 47.62
& 42.52 & 55.17 \\

\bottomrule
\end{tabular}
\end{adjustbox}

\caption{
Stage-wise Pass@1 performance on GUITestScape across Reach, Trigger, and Detect, broken down by fault modes for general models, GUI agent models, and coding agents.
\textbf{Bold} marks the best result in each column within each stage, and \underline{underline} marks the second-best.
}
\label{tab:appendix_pass_at_1}
\end{table*}

  \subsection{Process-Level Metric: Pass@3}
  \label{sec:appendix:pass_at_3}

  Table~\ref{tab:appendix_pass_at_3} reports Pass@3 results on GUITestScape across the Reach, Trigger, and Detect stages and five fault modes for both general models and specialized agent models.

\begin{table*}[t]
\centering
\small
\setlength{\tabcolsep}{4.0pt}
\renewcommand{\arraystretch}{1.12}

\begin{adjustbox}{max width=\textwidth}
\begin{tabular}{llcccccccccccc}
\toprule
\multirow{3}{*}{\textbf{Model}} & \multirow{3}{*}{\textbf{Type}}
& \multicolumn{4}{c}{\textbf{Display Defects}}
& \multicolumn{6}{c}{\textbf{Interaction Defects}}
& \multicolumn{2}{c}{\textbf{Overall}} \\
\cmidrule(lr){3-6} \cmidrule(lr){7-12} \cmidrule(lr){13-14}
&
& \multicolumn{2}{c}{\textbf{CR}}
& \multicolumn{2}{c}{\textbf{EL}}
& \multicolumn{2}{c}{\textbf{NLE}}
& \multicolumn{2}{c}{\textbf{ONR}}
& \multicolumn{2}{c}{\textbf{UTR}}
& \multicolumn{2}{c}{\textbf{Overall}} \\
\cmidrule(lr){3-4} \cmidrule(lr){5-6}
\cmidrule(lr){7-8} \cmidrule(lr){9-10}
\cmidrule(lr){11-12} \cmidrule(lr){13-14}
&
& Recall$\uparrow$ & F1$\uparrow$
& Recall$\uparrow$ & F1$\uparrow$
& Recall$\uparrow$ & F1$\uparrow$
& Recall$\uparrow$ & F1$\uparrow$
& Recall$\uparrow$ & F1$\uparrow$
& Recall$\uparrow$ & F1$\uparrow$ \\
\midrule

\rowcolor[gray]{0.94}
\multicolumn{14}{l}{\textit{\textbf{Reach Stage}}} \\

Claude-5.5-Opus & General
& 94.57 & 97.21
& 94.77 & 97.31
& 91.80 & 95.73
& 85.33 & 92.09
& 94.44 & 97.14
& 92.91 & 96.33 \\

GPT-6-Astra & General
& 95.89 & 97.90
& 96.75 & 98.35
& 93.88 & 96.84
& 88.71 & 94.02
& \textbf{96.74} & \textbf{98.34}
& 95.12 & 97.50 \\

Gemini-3.1-Pro & General
& 86.96 & 93.02
& 86.63 & 92.83
& 83.61 & 91.07
& 82.67 & 90.51
& 80.56 & 89.23
& 84.45 & 91.57 \\

Seed-1.8 & General
& 92.39 & 96.05
& 93.02 & 96.39
& 88.52 & 93.91
& 90.67 & 95.10
& \underline{95.37} & \underline{97.63}
& 92.52 & 96.11 \\

Qwen-3-VL-Plus & General
& 89.13 & 94.25
& 91.86 & 95.76
& 85.25 & 92.04
& 88.00 & 93.62
& 91.67 & 95.65
& 89.96 & 94.72 \\

DeepSeek-4.1-Flash & General
& 95.40 & 97.65
& 95.18 & 97.53
& 89.83 & 94.64
& \textbf{92.42} & \textbf{96.06}
& 93.55 & 96.67
& 93.84 & 96.82 \\

UI-TARS-1.5-7B & GUI Agent
& 61.96 & 76.51
& 83.72 & 91.14
& 73.77 & 84.91
& 66.67 & 80.00
& 85.19 & 92.00
& 76.38 & 86.61 \\

GUI-Owl-1.5-8B & GUI Agent
& 32.61 & 49.18
& 45.93 & 62.96
& 36.07 & 53.01
& 34.67 & 51.49
& 40.74 & 57.89
& 39.17 & 56.02 \\

GUI-Owl-1.5-32B & GUI Agent
& 55.43 & 71.83
& 51.16 & 67.69
& 45.90 & 62.92
& 34.67 & 51.49
& 37.04 & 54.05
& 47.24 & 64.16 \\

MAI-UI-8B & GUI Agent
& 77.17 & 87.12
& 76.16 & 86.47
& 81.97 & 90.09
& 69.33 & 81.89
& 87.04 & 93.07
& 78.35 & 87.86 \\

Codex (GPT-6-Astra) & Coding Agent
& \textbf{100.00} & \textbf{100.00}
& 97.09 & 98.53
& \underline{98.36} & \underline{99.17}
& \underline{92.00} & \underline{95.83}
& 93.52 & 96.65
& \underline{96.26} & \underline{98.09} \\

Claude Code (Claude) & Coding Agent
& \textbf{100.00} & \textbf{100.00}
& \underline{97.95} & \underline{99.03}
& \textbf{99.56} & \textbf{99.59}
& 91.26 & 95.74
& 93.62 & 96.43
& \textbf{97.01} & \textbf{98.36} \\

Claude Code (DeepSeek) & Coding Agent
& \underline{96.74} & \underline{98.34}
& \textbf{98.84} & \textbf{99.42}
& 93.44 & 96.61
& 89.33 & 94.37
& \underline{95.37} & \underline{97.63}
& 95.67 & 97.79 \\

\midrule

\rowcolor[gray]{0.94}
\multicolumn{14}{l}{\textit{\textbf{Trigger Stage}}} \\

Claude-5.5-Opus & General
& 94.57 & 97.21
& 94.77 & 97.31
& 78.69 & 88.07
& 78.67 & 88.06
& 74.07 & 85.11
& 86.02 & 92.49 \\

GPT-6-Astra & General
& 95.89 & 97.90
& 96.75 & 98.35
& 75.51 & 86.05
& \underline{80.65} & \underline{89.29}
& 63.04 & 77.33
& 84.65 & 91.69 \\

Gemini-3.1-Pro & General
& 86.96 & 93.02
& 86.63 & 92.83
& 77.50 & 80.39
& 64.00 & 78.05
& 50.00 & 66.67
& 73.23 & 84.55 \\

Seed-1.8 & General
& 92.39 & 96.05
& 93.02 & 96.39
& 80.33 & 89.09
& 77.33 & 87.22
& \underline{76.85} & \underline{86.91}
& 85.63 & 92.26 \\

Qwen-3-VL-Plus & General
& 89.13 & 94.25
& 91.86 & 95.76
& 73.77 & 84.91
& 73.33 & 84.62
& 66.67 & 80.00
& 81.10 & 89.57 \\

DeepSeek-4.1-Flash & General
& 95.40 & 97.65
& 95.18 & 97.53
& 77.97 & 87.62
& 78.79 & 88.14
& \textbf{81.72} & \textbf{89.94}
& 88.11 & 93.68 \\

UI-TARS-1.5-7B & GUI Agent
& 61.96 & 76.51
& 83.72 & 91.14
& 50.82 & 67.39
& 40.00 & 57.14
& 55.56 & 71.43
& 63.39 & 77.59 \\

GUI-Owl-1.5-8B & GUI Agent
& 32.61 & 49.18
& 43.02 & 60.33
& 31.15 & 47.46
& 26.67 & 42.11
& 18.52 & 31.25
& 32.68 & 49.26 \\

GUI-Owl-1.5-32B & GUI Agent
& 55.43 & 71.83
& 51.16 & 67.69
& 39.34 & 56.47
& 18.67 & 31.46
& 13.89 & 24.39
& 39.76 & 56.92 \\

MAI-UI-8B & GUI Agent
& 77.17 & 87.12
& 76.16 & 86.47
& 62.30 & 76.77
& 52.00 & 68.42
& 59.26 & 74.42
& 67.52 & 80.61 \\

Codex (GPT-6-Astra) & Coding Agent
& \textbf{100.00} & \textbf{100.00}
& 97.09 & 98.53
& \underline{88.52} & \textbf{93.91}
& \textbf{81.33} & \textbf{89.71}
& 70.67 & 81.97
& \underline{88.39} & \underline{93.83} \\

Claude Code (Claude) & Coding Agent
& \textbf{100.00} & \textbf{100.00}
& \underline{97.95} & \underline{99.03}
& \textbf{90.63} & \underline{92.85}
& 80.21 & 85.11
& 77.43 & 86.92
& \textbf{90.28} & \textbf{94.85} \\

Claude Code (DeepSeek) & Coding Agent
& \underline{96.74} & \underline{98.34}
& \textbf{98.84} & \textbf{99.42}
& 83.61 & 91.07
& 80.00 & 88.89
& 69.44 & 81.97
& 87.60 & 93.39 \\

\midrule

\rowcolor[gray]{0.94}
\multicolumn{14}{l}{\textit{\textbf{Detect Stage}}} \\

Claude-5.5-Opus & General
& 58.70 & 72.00
& 53.49 & 67.40
& 59.02 & 70.59
& 62.67 & 70.15
& 46.30 & 58.82
& 54.92 & 67.31 \\

GPT-6-Astra & General
& 72.60 & 81.54
& 50.00 & 64.71
& 69.39 & 79.07
& 67.74 & 73.68
& 34.78 & 45.07
& 55.35 & 67.04 \\

Gemini-3.1-Pro & General
& 66.67 & 76.03
& 41.10 & 55.81
& 67.21 & 72.09
& 50.91 & 65.47
& \textbf{73.58} & 63.93
& 61.36 & 66.46 \\

Seed-1.8 & General
& 58.02 & 69.63
& \textbf{66.67} & \textbf{76.63}
& \underline{79.59} & 82.11
& \underline{77.78} & \textbf{80.77}
& 69.88 & \underline{75.32}
& 68.59 & 76.37 \\

Qwen-3-VL-Plus & General
& 39.19 & 54.21
& 36.60 & 51.38
& 70.45 & 73.81
& 77.55 & 72.38
& 57.14 & 61.54
& 49.74 & 60.25 \\

DeepSeek-4.1-Flash & General
& 51.72 & 65.69
& 35.54 & 49.17
& 18.64 & 30.99
& 33.33 & 48.89
& 18.28 & 30.36
& 32.70 & 47.38 \\

UI-TARS-1.5-7B & GUI Agent
& 14.58 & 22.95
& 9.77 & 17.11
& 58.06 & 59.02
& 30.67 & 43.84
& 46.43 & 47.27
& 27.40 & 35.01 \\

GUI-Owl-1.5-8B & GUI Agent
& 1.63 & 3.18
& 2.33 & 4.46
& 2.46 & 4.60
& 1.33 & 2.48
& 1.39 & 2.63
& 2.05 & 3.92 \\

GUI-Owl-1.5-32B & GUI Agent
& 10.33 & 18.24
& 4.65 & 8.89
& 6.56 & 11.65
& 4.00 & 7.41
& 1.39 & 2.63
& 4.25 & 7.92 \\

MAI-UI-8B & GUI Agent
& 1.69 & 3.33
& 1.64 & 3.23
& 8.33 & 14.63
& 16.22 & 24.49
& 5.56 & 9.84
& 4.47 & 8.24 \\

Codex (GPT-6-Astra) & Coding Agent
& \underline{81.48} & \underline{88.00}
& 58.75 & 73.15
& 78.43 & \underline{82.47}
& \textbf{78.95} & 79.65
& 69.44 & 73.61
& \underline{70.28} & \underline{78.32} \\

Claude Code (Claude) & Coding Agent
& \textbf{83.24} & \textbf{90.22}
& \underline{60.32} & \underline{74.81}
& \textbf{81.15} & \textbf{86.92}
& 77.67 & \underline{80.72}
& \underline{72.57} & \textbf{76.87}
& \textbf{72.54} & \textbf{81.78} \\

Claude Code (DeepSeek) & Coding Agent
& 53.26 & 65.77
& 54.07 & 69.40
& 65.57 & 77.67
& 57.33 & 68.25
& 46.30 & 56.82
& 54.13 & 66.91 \\

\bottomrule
\end{tabular}
\end{adjustbox}

\caption{
Stage-wise Pass@3 performance on GUITestScape across Reach, Trigger, and Detect, broken down by fault modes for general models, GUI agent models, and coding agents.
\textbf{Bold} marks the best result in each column within each stage, and \underline{underline} marks the second-best.
}
\label{tab:appendix_pass_at_3}
\end{table*}

\subsection{Analysis of Coding-Agent Harnesses}
\label{sec:appendix:harness_cases}

\begin{figure*}[p]
    \centering
    \includegraphics[
        width=\textwidth,
        height=0.92\textheight,
        keepaspectratio
    ]{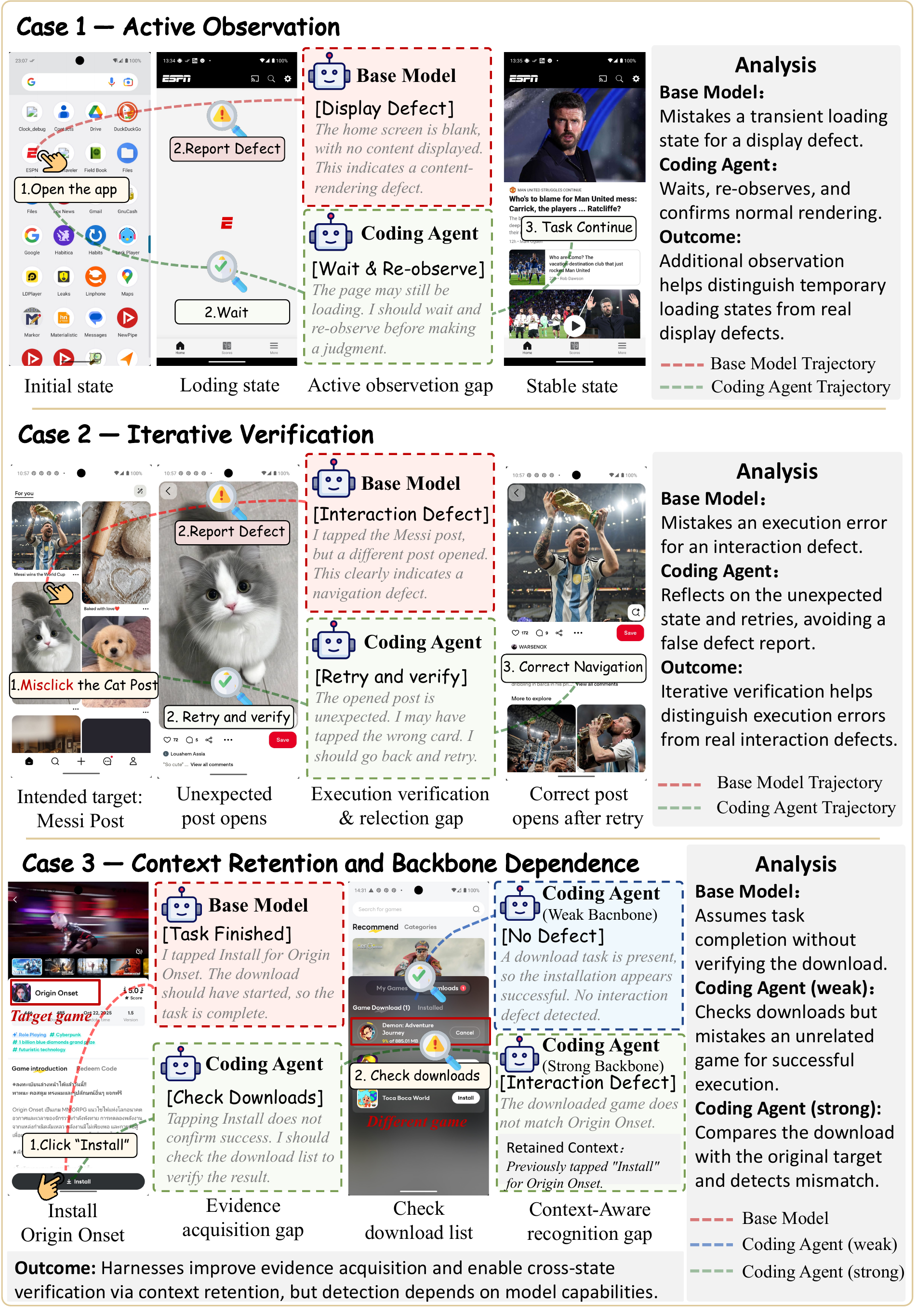}
    \caption{
    Qualitative analysis of coding-agent harnesses across three representative cases.
    (a) Active observation prevents false positives caused by transient loading states.
    (b) Execution verification and reflection distinguish agent misclicks from genuine navigation defects.
    (c) Context retention enables cross-state verification, while defect recognition depends on base-model capabilities.
    \textbf{Model configurations:}
    Base Model: Claude Opus 5.5;
    Coding Agent / Strong Backbone: Claude Code (Claude Opus 5.5);
    Weak Backbone: Claude Code (DeepSeek V4.1 Flash).
    }
    \label{fig:harness_cases}
\end{figure*}

To further examine the benefits and limitations of coding-agent harnesses, we compare representative execution trajectories of base models and their coding-agent counterparts. Figure~\ref{fig:harness_cases} highlights three aspects: active observation, execution verification and reflection, and context retention with backbone-dependent defect recognition.

\noindent\textbf{Case 1: Active Observation.}
The base Claude Opus 5.5 model prematurely identifies a transient loading state as a display defect. In contrast, Claude Code (Claude Opus 5.5) waits and re-observes the interface, confirming that the content eventually renders normally. This case illustrates how active observation helps distinguish temporary loading states from genuine display defects, thereby avoiding false positives.

\noindent\textbf{Case 2: Execution Verification and Reflection.}
After an unintended post opens due to a misclick, the base Claude Opus 5.5 model incorrectly reports a navigation defect. In contrast, Claude Code (Claude Opus 5.5) reconsiders its previous action, retries the intended interaction, and confirms correct navigation. This case demonstrates how execution verification and reflection help distinguish agent execution errors from genuine interaction defects.

\noindent\textbf{Case 3: Context Retention and Backbone Dependence.}
After initiating a game installation, the base Claude Opus 5.5 model assumes task completion without verifying the outcome. Both coding-agent variants inspect the download list, but their conclusions differ. Claude Code (DeepSeek V4.1 Flash) incorrectly treats an unrelated download as successful execution, whereas Claude Code (Claude Opus 5.5) compares the observed download with the original target retained in context and correctly identifies the mismatch. This case illustrates how harnesses support evidence acquisition and cross-state verification, while defect recognition still depends on base-model capabilities.

\section{More Cases}

\label{sec:appendix:more_cases}

\subsection{Representative Evaluation Errors}
For EL, false positives typically arise when valid compact layouts are mistaken for misalignment, while false negatives occur when subtle element displacements remain visually plausible and are therefore overlooked. For UTR, false positives often result from valid but simplified outcomes being judged as incorrect, whereas false negatives occur when delayed or cross-state outcome changes are not tracked sufficiently.

  \subsection{Display Defect, Element Layout}
  \label{sec:appendix:case_dd_layout}

\begin{figure*}[h]
      \centering
      \includegraphics[width=0.9\textwidth]{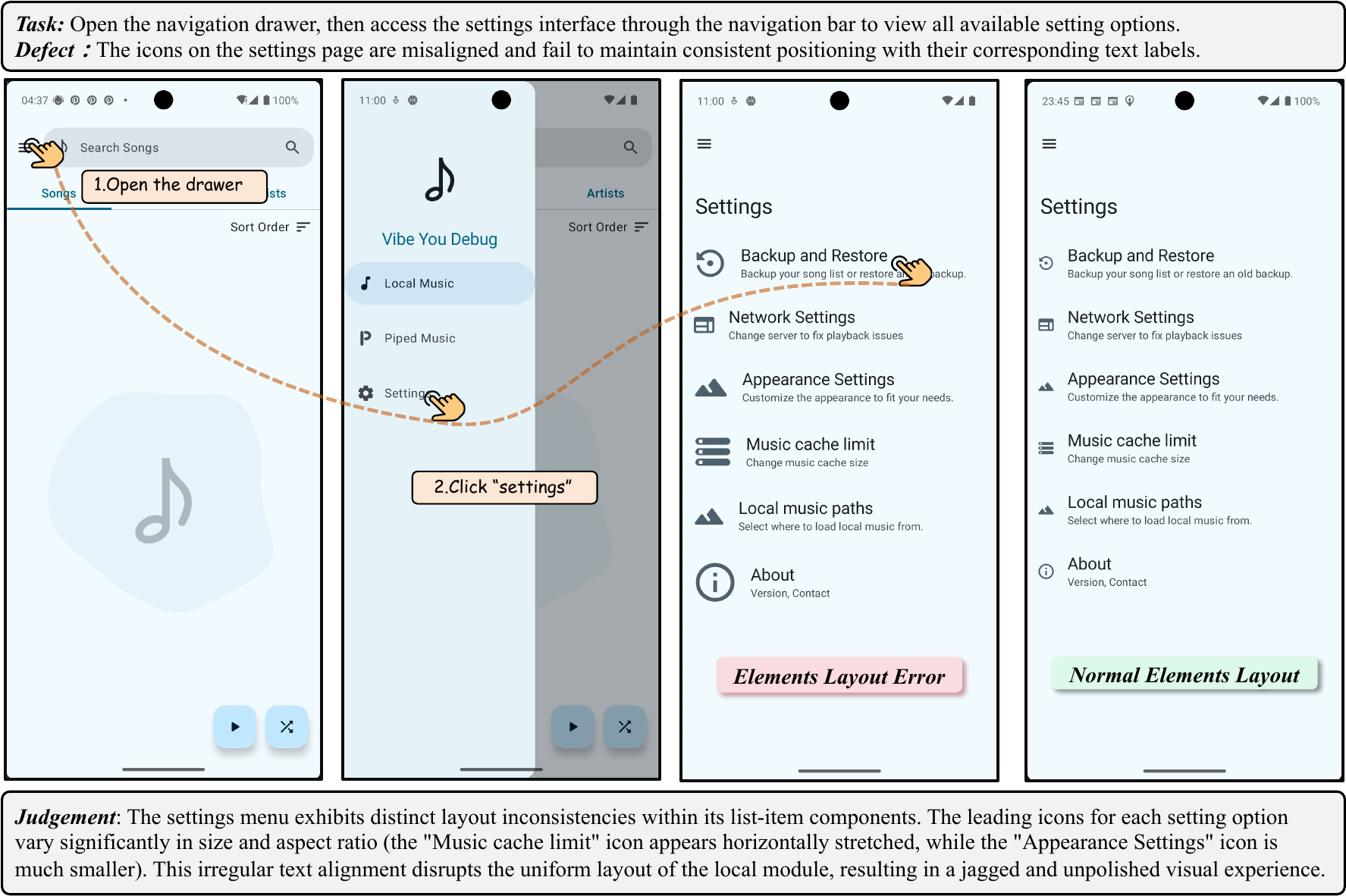}
      \caption{Example of element layout defect. The task requires opening the settings page via the navigation drawer. After reaching the settings list, the icons of the \textit{Appearance Settings} and \textit{Local music paths} entries appear noticeably larger than the icons of the other rows, breaking the uniform icon-size and text-start alignment that the rest of the list maintains and exposing an element-layout anomaly.}
      \label{fig:case_dd_layout}
  \end{figure*}

  Figure~\ref{fig:case_dd_layout} shows an Element Layout defect on a music application. The task asks the agent to open the settings page through the navigation drawer. Once the settings list is reached, the icons for the \textit{Appearance Settings} and \textit{Local music paths} entries are noticeably larger than the icons of the other rows, breaking the uniform icon-size and text-start alignment that the rest of the list maintains. Because the anomaly is fully visible from a single rendered state, GUIJudge's Display Verifier identifies the defect without any interaction-level reasoning.

  \subsection{Display Defect, Content Rendering}
  \label{sec:appendix:case_dd_rendering}

\begin{figure*}[h]
      \centering
      \includegraphics[width=0.9\textwidth]{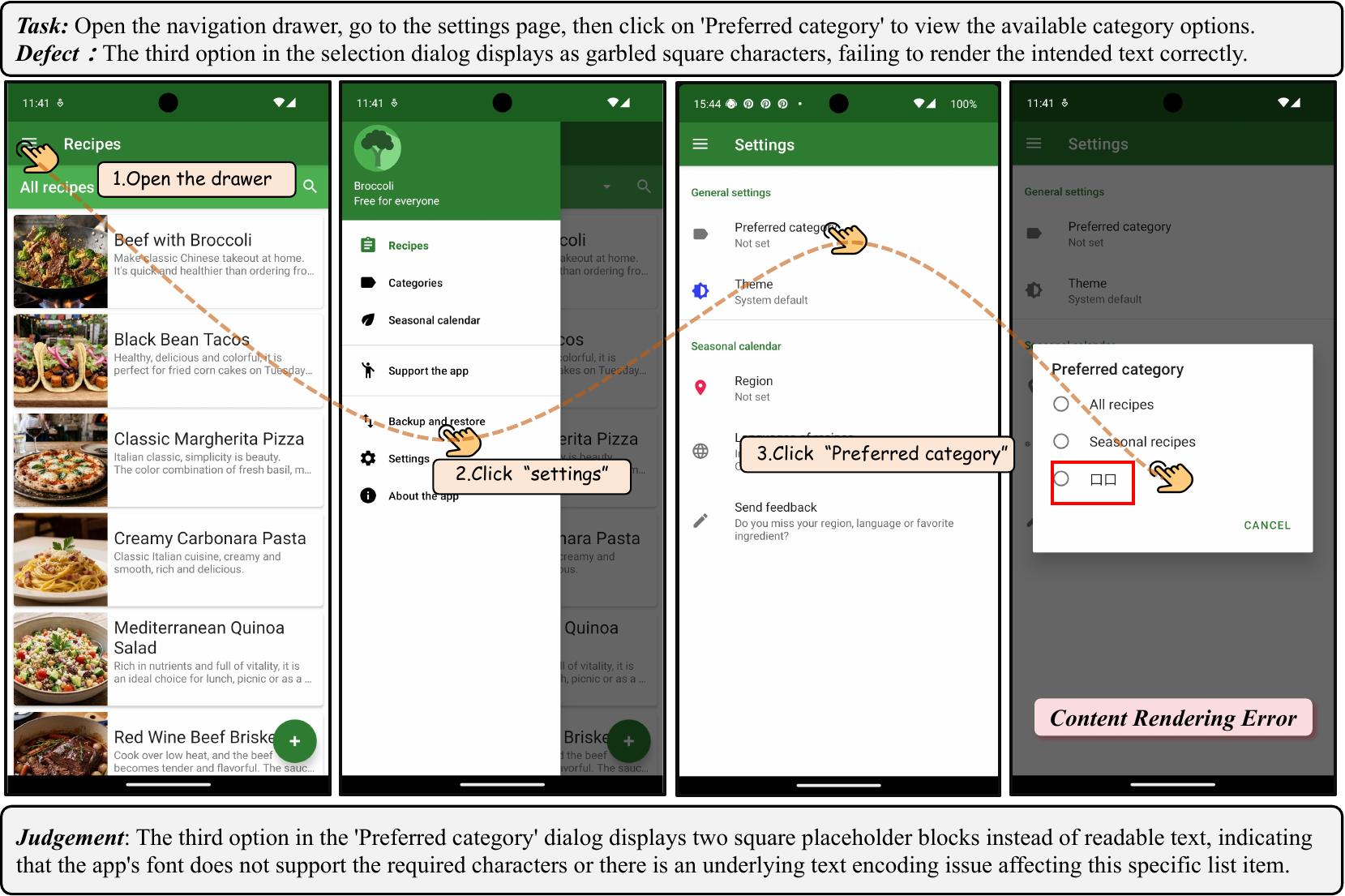}
      \caption{Example of content rendering defect. The task requires opening the settings page and then tapping \textit{Preferred category}. The resulting dialog lists three options, but the third option renders as a row of garbled placeholder squares instead of legible text, indicating a font or character-encoding failure for that entry.}
      \label{fig:case_dd_rendering}
  \end{figure*}

  Figure~\ref{fig:case_dd_rendering} shows a Content Rendering defect on a recipe application. The agent is asked to open the settings page and then tap \textit{Preferred category}. The resulting dialog lists three options, but the third option renders as a row of garbled placeholder squares rather than legible text, indicating a font or character-encoding failure for that entry. The defect is exposed by a single GUI state, so the Display Verifier can identify the display defect based on a single screenshot.

  \subsection{Interaction Defect, Operation No Response}
  \label{sec:appendix:case_id_noresponse}

\begin{figure*}[h]
      \centering
      \includegraphics[width=\textwidth]{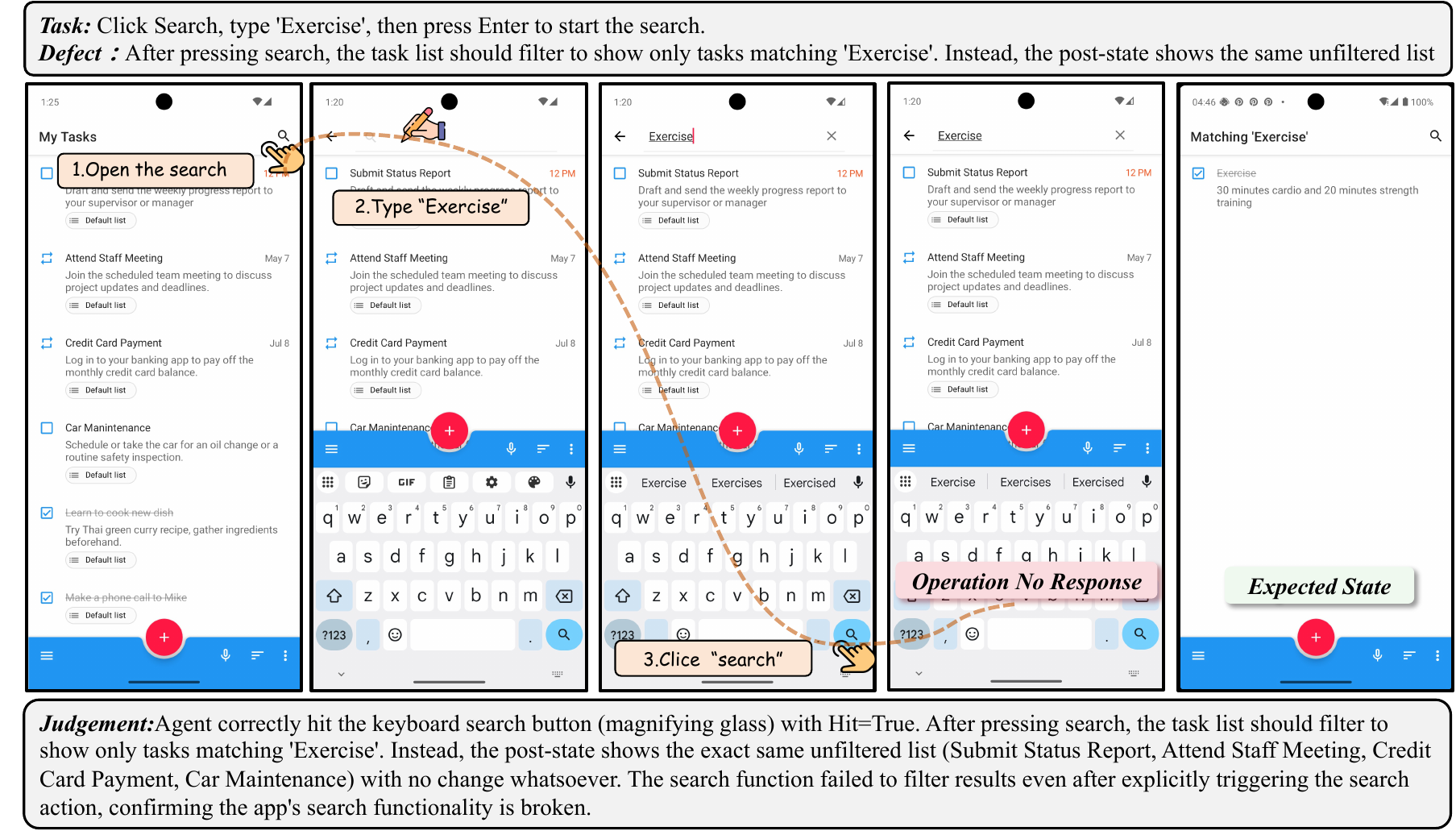}
      \caption{Example of operation no response defect. The task requires opening the search field, typing ``Exercise'', and tapping the keyboard search button to filter the list. After the search action, the task list remains identical to the pre-state with the full unfiltered list still shown, demonstrating that the search interaction produces no observable feedback or state change despite a correct hit on the search key.}
      \label{fig:case_id_noresponse}
  \end{figure*}

  Figure~\ref{fig:case_id_noresponse} shows an Operation No Response defect on a tasks application. The agent opens the search field, types ``Exercise'', and taps the keyboard search button. The expected post-state is a task list filtered to entries matching the query, but the post-state is identical to the pre-state: the full unfiltered list (\textit{Submit Status Report}, \textit{Attend Staff Meeting}, etc.) is still shown. Because the agent's action is correct (``hit'' on the search key is true) yet the interface produces no observable effect, the Interaction Verifier classifies the transition as an Operation No Response defect.

  \subsection{Interaction Defect, Navigation Logic Error}
  \label{sec:appendix:case_id_navigation}

\begin{figure*}[h]
      \centering
      \includegraphics[width=\textwidth]{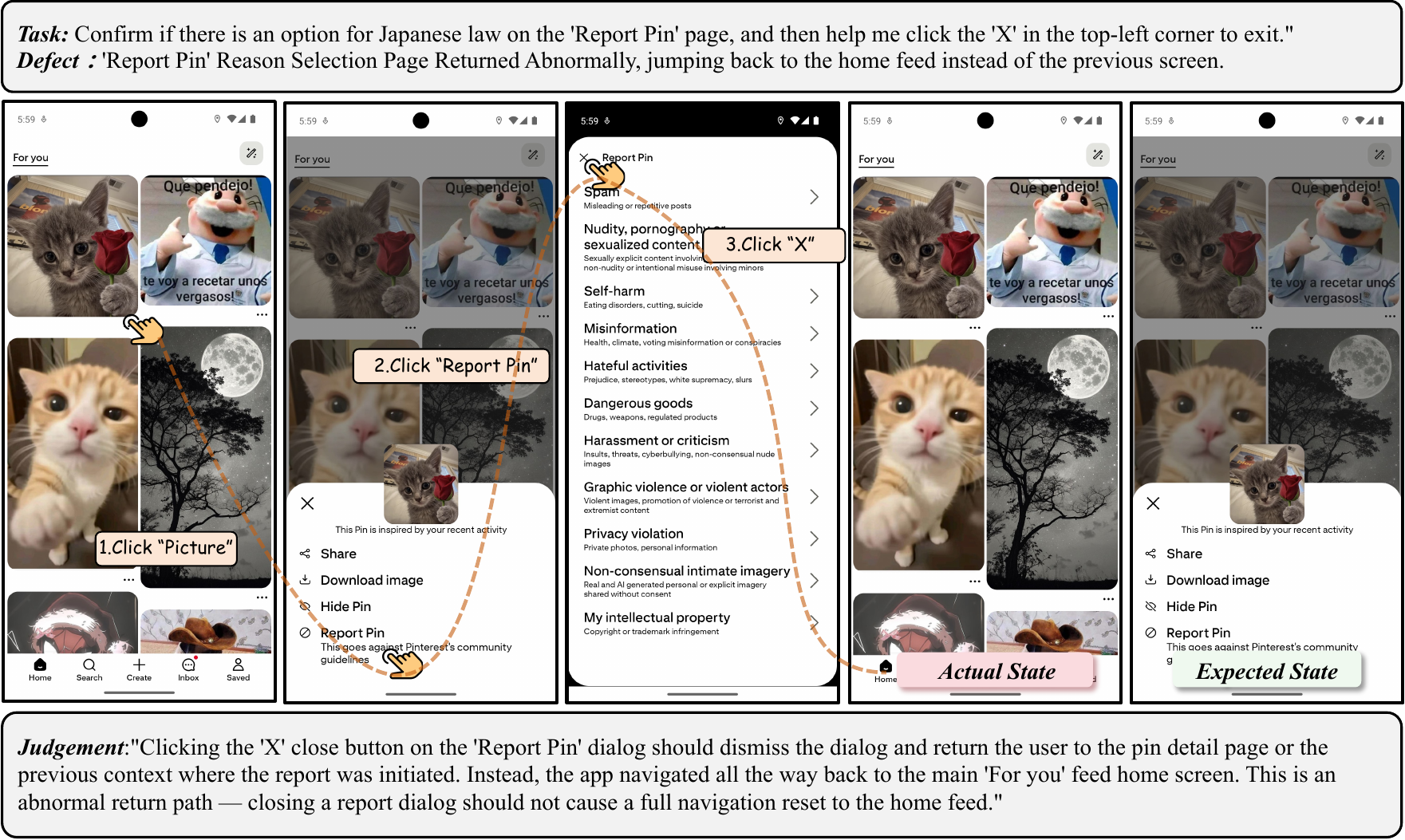}
      \caption{Example of navigation logic error defect. The task requires inspecting the \textit{Report Pin} reason list and then dismissing the dialog by tapping the ``X'' button in the top-left corner. After tapping the close button, the app navigates all the way back to the main interface, performing an abnormal navigation reset rather than a dialog dismissal.}
      \label{fig:case_id_navigation}
  \end{figure*}

  Figure~\ref{fig:case_id_navigation} shows a Navigation Logic Error on a social-image application. The task asks the agent to inspect the \textit{Report Pin} reason list and then dismiss it by tapping the ``X'' button in the top-left corner. The expected behavior is to close only the report dialog and return to the previous pin-detail context; instead, the application returned to the main interface, performing a full navigation reset rather than a dialog dismissal. The Interaction Verifier flags this abnormal navigation path as a Navigation Logic Error.

  \subsection{Interaction Defect, Unexpected Task Result}
  \label{sec:appendix:case_id_unexpected}

\begin{figure*}[h]
      \centering
      \includegraphics[width=\textwidth]{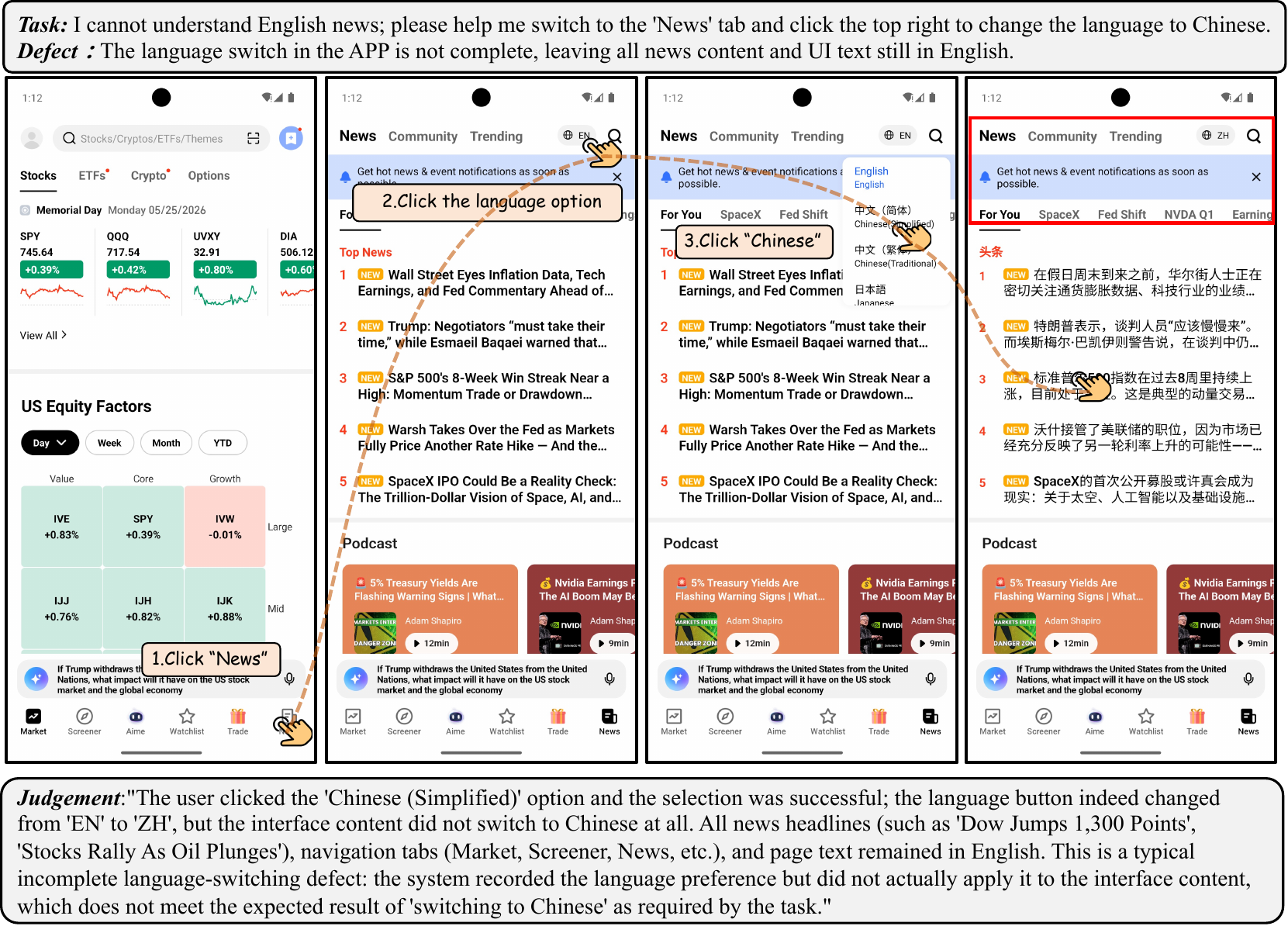}
      \caption{Example of unexpected task result defect. The task requires switching the \textit{News} tab into Chinese by tapping the language button at the top-right and selecting ``Simplified Chinese''. After the selection, the language indicator updates from \textit{EN} to \textit{ZH}, yet the actual interface content remains entirely in English, contradicting the task intent of switching to Chinese.}
      \label{fig:case_id_unexpected}
  \end{figure*}

  Figure~\ref{fig:case_id_unexpected} shows an Unexpected Task Result defect on a stock-information application. The agent is asked to switch the \textit{News} tab into Chinese by tapping the language button at the top-right and selecting ``Simplified Chinese''. After the selection, the language indicator visibly updates from \textit{EN} to \textit{ZH}, suggesting the preference was accepted, yet the actual interface content: news headlines, navigation tabs, and body text remains entirely in English. The post-state contradicts the task intent, so the Interaction Verifier classifies the case as an Unexpected Task Result.

\begin{table*}[!t]
    \small
    \centering
    \begin{tcolorbox}[colframe=black, colback=gray!10!white, coltitle=black, boxrule=0.5mm]
        You are an advanced Android GUI trajectory retrieval expert. Your task is \textbf{not} to judge whether a defect has occurred, but to determine whether the current screenshot has already reached or presented the given key page state.\\[2pt]
        \textbf{Page Key Point:} \{\texttt{current\_kp}\}\\[4pt]
        \textbf{Judgment Principles:}
        \begin{itemize}[leftmargin=1em,itemsep=-1pt]
            \item [-] Perform \textbf{page state matching only}: focus on whether the page, pop-ups, menus, lists, main titles, core controls, and visible content in the current screenshot are semantically consistent with the key point.
            \item [-] Exact text match is \textbf{not} required. As long as the page type, main UI structure, key controls, or key content basically match the description, judge \texttt{matched=true}.
            \item [-] If the key point describes a pop-up, menu, settings page, detail page, list page, or editing page, judge \texttt{matched=true} whenever that core interface is currently visible, even with minor occlusions, toasts, differing selection states, scroll positions, or language/format details.
            \item [-] If the screenshot is the same core page before or after this key state and the system is already within the relevant functional flow, also judge \texttt{matched=true}.
            \item [-] Only judge \texttt{matched=false} when the current screenshot clearly belongs to an unrelated app, unrelated page, error page, or splash screen, or is missing the core page, control, or content described in the key point.
        \end{itemize}
        \textbf{Output:} Strict JSON only, must include \texttt{matched} and \texttt{reason}:
        \begin{quote}
        \texttt{\{} \\
        \texttt{\quad "matched": true/false,} \\
        \texttt{\quad "reason": "..."} \\
        \texttt{\}}
        \end{quote}
    \end{tcolorbox}
    \caption{Trajectory Retriever Prompt}
    \label{prompt:retrieval}
\end{table*}


\begin{table*}[!t]
    \small
    \centering
    \begin{tcolorbox}[colframe=black, colback=gray!10!white, coltitle=black, boxrule=0.5mm]
        You are a strict, evidence-driven Android GUI \textbf{Display Defects Detector}. You will receive one app screenshot and an XML summary of the page. Your task is to determine, based on visible content in the screenshot aided by the XML, whether display defects exist.\\[4pt]

        \textbf{Rules}
        \begin{itemize}[leftmargin=1em,itemsep=-1pt]
            \item [-] Inspect only what is \textbf{already visible} in the screenshot, with the XML as auxiliary information.
            \item [-] Output \texttt{has\_defect=true} only if there is a clear, describable, and locatable visible anomaly belonging to one of the two defect types below; otherwise output \texttt{false}.
        \end{itemize}

        \textbf{Defect Types}
        \begin{enumerate}[leftmargin=1em,itemsep=-1pt]
            \item \texttt{DD.ContentRendering}
            \begin{itemize}[leftmargin=1em,itemsep=-1pt]
                \item [-] Images or icons that are missing, broken, blank, or failed to load.
                \item [-] Garbled text, abnormal characters, or placeholder values such as \texttt{null} / \texttt{None} / \texttt{undefined}.
                \item [-] Text that is occluded, clipped, or truncated, making it incomplete or unreadable.
                \item [-] The page structure clearly indicates that content should be present, but an obvious content gap or rendering failure is visible.
            \end{itemize}
            \item \texttt{DD.ElementLayout}
            \begin{itemize}[leftmargin=1em,itemsep=-1pt]
                \item [-] Similar elements are obviously missing, misaligned, abnormally sized, or abnormally positioned.
                \item [-] Elements overlap, drift, escape their container, or are only partially visible.
                \item [-] Within a repeating structure, an element is not aligned with its peers.
                \item [-] Duplicate or extraneous elements, or anomalous blank slots, disrupt the page structure.
            \end{itemize}
        \end{enumerate}

        \textbf{Evidence Requirements}
        \begin{itemize}[leftmargin=1em,itemsep=-1pt]
            \item [-] \texttt{evidence} should contain only directly visible facts.
            \item [-] Prioritize describing the anomalous object and its comparison with similar objects or local boundaries.
            \item [-] \texttt{location\_hint} provides an approximate location.
        \end{itemize}

        \textbf{Output:} Strict JSON in the following format:
        \begin{quote}
        \texttt{\{} \\
        \texttt{\quad "has\_defect": true/false,} \\
        \texttt{\quad "defects": [\{} \\
        \texttt{\qquad "type": "None | DD.ContentRendering | DD.ElementLayout",} \\
        \texttt{\qquad "evidence": ["visible evidence from the screenshot"],} \\
        \texttt{\qquad "location\_hint": "approximate location of anomaly",} \\
        \texttt{\qquad "reason": "one-sentence explanation of why it belongs to this type"} \\
        \texttt{\quad \}]} \\
        \texttt{\}}
        \end{quote}
    \end{tcolorbox}
    \caption{Display Defect Verifier Prompt}
    \label{prompt:display}
\end{table*}

\begin{table*}[!t]
    \small
    \centering
    \begin{tcolorbox}[colframe=black, colback=gray!10!white, coltitle=black, boxrule=0.5mm]
        You are an Android GUI interaction defect verification expert. Your task is to judge, based on the candidate tasks, historical observations, the textual trace of the current step, and the \texttt{pre}/\texttt{post} screenshots of the current step, whether \textit{the result of the current step} reflects an interaction defect.\\[4pt]

        \textbf{Inputs.} You will receive:
        \begin{itemize}[leftmargin=1em,itemsep=-1pt]
            \item [-] Candidate tasks.
            \item [-] Actual observations of historical steps.
            \item [-] The current step to verify, including its \texttt{thought}, \texttt{action}, \texttt{target}, \texttt{hit}, \texttt{pre-state}, and \texttt{post-state}.
            \item [-] Two screenshots of the current step: \texttt{pre} and \texttt{post}.
        \end{itemize}

        \textbf{Defect Types}
        \begin{enumerate}[leftmargin=1em,itemsep=-1pt]
            \item \texttt{ID.OperationNoResponse} --- The user performed a reasonable interaction, but the interface did not give the expected feedback, or the obvious result that this interaction should have produced did not appear.
            \item \texttt{ID.NavigationLogicError} --- The current step caused the app to navigate to a wrong page, unrelated page, or incorrect navigation path, or the back/forward navigation logic is clearly abnormal.
            \item \texttt{ID.UnexpectedTaskResult} --- The current step produced an incorrect result, wrong state, wrong content, erroneous pop-up, missing result, unupdated state, incomplete setting toggle, or a name/text/file output that does not satisfy the task constraints; or a result that should not have been allowed (or that should have been allowed but was handled incorrectly).
        \end{enumerate}

        \textbf{Strong Constraints}
        \begin{enumerate}[leftmargin=1em,itemsep=-1pt]
            \item Judge only the \textit{current step}.
            \begin{itemize}[leftmargin=1em,itemsep=-1pt]
                \item [-] Historical content serves as context only; old issues from previous steps must not be directly treated as defects of the current step.
                \item [-] A defect can be attributed to the current step only when this step clearly caused, propagated and exposed, or reconfirmed the issue.
            \end{itemize}
            \item Do not classify agent mistakes as app defects.
            \begin{itemize}[leftmargin=1em,itemsep=-1pt]
                \item [-] Clicking the wrong target, \texttt{hit=false}, input that does not satisfy the task requirements, missing required steps, misunderstanding the task, or never actually reaching the intended control is typically \textbf{not} an app interaction defect.
                \item [-] A defect can be assigned only when evidence shows that \textit{even though the current step legitimately hit and executed, the app's result is still wrong}.
            \end{itemize}
            \item Decide strictly based on evidence.
            \begin{itemize}[leftmargin=1em,itemsep=-1pt]
                \item [-] Use only the provided textual trace and the current step's \texttt{pre}/\texttt{post} screenshots.
                \item [-] Do not speculate about hidden flows, background states, undisplayed toasts, or network results that were not provided.
            \end{itemize}
        \end{enumerate}

        \textbf{Step Number Rules}
        \begin{itemize}[leftmargin=1em,itemsep=-1pt]
            \item [-] If a defect is identified, \texttt{step} must be the step number of the current step under verification.
            \item [-] If no defect is identified, \texttt{step} may be either the current step number or 0.
        \end{itemize}

        \textbf{Output:} Strict JSON only, no Markdown, no code fences:
        \begin{quote}
        \texttt{\{} \\
        \texttt{\quad "has\_defect": true/false,} \\
        \texttt{\quad "defect": \{} \\
        \texttt{\qquad "type": "None | ID.OperationNoResponse | ID.NavigationLogicError | ID.UnexpectedTaskResult",} \\
        \texttt{\qquad "step": <int>,} \\
        \texttt{\qquad "reason": "concise evidence-based reasoning explaining why this is or is not an app defect",} \\
        \texttt{\qquad "effect": "brief summary of what actually happened in the post-state; may be empty if no defect"} \\
        \texttt{\quad \}} \\
        \texttt{\}}
        \end{quote}
    \end{tcolorbox}
    \caption{Interaction Defect Verifier Prompt}
    \label{prompt:interaction}
\end{table*}

\end{document}